\documentclass[]{aastex631}
\usepackage{graphicx,amsmath,amssymb,amstext}
\usepackage{amssymb,amsbsy,amsfonts,amsthm,color}

\usepackage{mathtools,nccmath}

\usepackage{amssymb}
\usepackage{amsmath}
\usepackage{comment}
\usepackage{graphicx}
\usepackage{tabularx}
\usepackage{color}

\usepackage{orcidlink}
\usepackage{lipsum}

\begin{document}

\title{Can a hybrid star with constant sound speed parametrization explain the  new NICER mass-radius measurements ?
}

\author{Suman Pal,\orcidlink{0009-0000-5944-4261}}
\email{sumanvecc@gmail.com}
\affiliation{Physics Group, Variable Energy Cyclotron Centre, 1/AF Bidhan Nagar, Kolkata 700064, India}
\affiliation{Homi Bhabha National Institute, Training School Complex, Anushakti Nagar, Mumbai 400085, India} 
\author{Gargi Chaudhuri,\orcidlink{0000-0002-8913-0658}}
\email{gargi@vecc.gov.in}
\affiliation{Physics Group, Variable Energy Cyclotron Centre, 1/AF Bidhan Nagar, Kolkata 700064, India}
\affiliation{Homi Bhabha National Institute, Training School Complex, Anushakti Nagar, Mumbai 400085, India}

\begin{abstract}

We present a reanalysis of NICER observations of PSR J0740+6620 and PSR J0030+0451  in order to test the consistency of various nuclear equations of state (EoS) within the framework of hybrid star models. In particular, we examine how different surface temperature models for PSR J0030+0451 -- categorized as Scenarios A, B, and C, lead to significantly different mass-radius estimates. 
 We present a comprehensive study constraining the parameters of the constant speed of sound model based on representative observational categories.
Our findings indicate that for certain hadronic equations of state(for representative case we chose one density independent and one density dependent )the results remain consistent for lower values of the energy density discontinuity, but discrepancies emerge as the discontinuity increases. Scenarios involving large jumps in energy density are generally disfavored by the requirement of supporting massive neutron stars, while higher values of the speed of sound in the quark matter phase tend to yield better agreement with various observational trends. These results underscore the importance of phase transition characteristics in aligning hybrid star models with current astrophysical observations.
We have also constrained the CSS parameters using observational data from PSR J0740+6620 and PSR J0952-0607 by computing the maximum mass supported by these parameters. 
\end{abstract}


\section{Introduction} 
\label{sect:intro} 

The Neutron Star Interior Composition Explorer (NICER), mounted on the International Space Station, is designed to observe the soft thermal X-ray emissions from rotation-powered millisecond pulsars (MSPs). These emissions, which originate from the heated polar caps of neutron stars, can be modeled to extract precise measurements of neutron star masses and radii. These measurements, in turn, provide important constraints on the equation of state (EoS) for cold, dense matter. To date, the NICER team has published detailed findings for two pulsars: PSR J0030+0451 \cite{Miller:2019cac,Riley:2019yda} and the massive PSR J0740+6620 \cite{Miller:2021qha,Riley:2021pdl,Salmi:2022cgy,Vinciguerra:2023qxq,Brandes:2024wpq,Li:2024sft}. NICER observes X-ray pulsations from MSPs caused by return currents heating the magnetic poles of neutron stars. These pulsations, shaped by relativistic effects, carry information about the surrounding spacetime. Using pulse profile modeling (PPM), which accounts for these effects, researchers can determine the star’s mass and radius, along with the size and position of the hot spots, 
 based on assumed models for their shape.
Observations of neutron stars have become one of the most effective ways to constrain the equation of state (EOS) of dense, strongly interacting matter in recent years. Given the strong link between the EOS and the mass–radius relationship of neutron stars, the most straightforward approach is to accurately measure the masses and radii of well-known neutron stars.

A recent reanalysis of data from the two-solar-mass millisecond pulsar PSR J0740+6620 and the 1.4 $M_{\odot}$ canonical pulsar PSR J0030+0451 poses a challenge to the theoretical description of the nuclear equation of state.
Recently, \cite{Li:2024sft} used reanalyzed data to test the predictions of EoS models that allow for strong first-order phase transitions, highlighting their impact on the neutron star mass-radius relation. Specifically, they examine three scenarios-labeled A, B, and C---that use the same reanalyzed data for PSR J0740+6620 but differ in their treatment of PSR J0030+0451, employing distinct models for the surface temperature distribution. Scenarios A, B, and C correspond to different surface temperature models: ST+PST, ST+PDT, and PDT-U, respectively. While all three adopt the same data for PSR J0740+6620, they vary in the modeling of PSR J0030+0451, leading to different inferences for the mass, radius, and likelihood values.
The inferred mass and radius values at the 1$\sigma$ level \cite{Vinciguerra:2023qxq} for PSR J0030+0451 differ across the three scenarios: for A (ST+PST), $M = 1.37^{+0.17}_{-0.17} \, M_\odot$, $R = 13.11^{+1.30}_{-1.30} \, \text{km}$; for B (ST+PDT), $M = 1.20^{+0.14}_{-0.11} \, M_\odot$, $R = 11.16^{+0.90}_{-0.80} \, \text{km}$; and for C (PDT-U), $M = 1.41^{+0.20}_{-0.19} \, M_\odot$, $R = 13.12^{+1.35}_{-1.21} \, \text{km}$.
The analysis by \cite{Vinciguerra:2023qxq}, based on Bayesian model comparison using the logarithm of the evidence $(\log \varepsilon)$, establishes a clear hierarchy of preference among the surface emission models. Specifically, the joint NICER+XMM analysis strongly favors the ‘PDT-U’ model (Scenario C), considers the ‘ST+PDT’ model (Scenario B) a less probable alternative, and decisively rules out the ‘ST+PST’ model (Scenario A) as  being inconsistent with the data.
In light of this, we do not treat all  the three scenarios (A, B, and C) to be  equally plausible. Our analysis  primarily focuses on  the scenarios C and B, which remain supported by current observations. Results corresponding to Scenario A are included for completeness but are clearly marked as statistically disfavored.
In addition to NICER measurements, gravitational wave observatories such as LIGO offer an independent avenue to constrain the neutron star equation of state (EOS) via tidal deformability. The gravitational wave event GW170817 \cite{LIGOScientific:2017vwq}, for instance, favors softer EOSs that yield more compact neutron star configurations.

The relativistic mean field (RMF) model \cite{Glendenning:1997wn,Sun:2023xkg,Dutra:2014qga} is a widely used framework for describing the equation of state (EoS) of neutron stars. At the extreme densities and pressures in their cores, matter may undergo a phase transition to deconfined quark matter, leading to the formation of a hybrid star \cite{Glendenning:1997wn,Blaschke:2020vuy,Weissenborn:2011qu,Bhattacharyya:2009fg,Gomes:2018eiv,Han:2019bub,Ferreira:2020evu, Most:2018eaw,Zha:2020gjw,Xia:2019pnq,Maslov:2018ghi,Montana:2018bkb,Christian:2018jyd,Lugones:2021tee,Bozzola:2019tit, Liu:2022mje}.---a neutron star with a quark core surrounded by hadronic matter.

To model the hadronic phase, we employ several RMF parameterizations with different interaction schemes, allowing us to investigate how hybrid star properties depend on the underlying nuclear physics.
The study of phase transitions in hybrid stars provides a unique opportunity to explore the behavior of the speed of sound ($C_s$) in dense matter. A physically consistent EoS must satisfy the condition $0 \leq C_s^2 \leq 1$, ensuring both causality and thermodynamic stability. However, whether a more stringent upper bound on $C_s^2$ exists in the high-density regime remains an open question---particularly in the context of deconfined quark matter.

In this work, our primary objective is to analyze reanalyzed pulsar data within the framework of hybrid stars. For this, we utilize a constant speed of sound (CSS) model to describe the high-density quark matter phase and investigate its impact on the structure of hybrid stars.
The flexibility of the CSS model makes it particularly suitable for studying the phase transition in hybrid stars, as it allows for the exploration of various parameter dependencies. Specifically, we examine the properties of hybrid stars by considering factors such as the speed of sound, energy gap, and transition density. 
In line with the statistical evidence presented by \cite{Vinciguerra:2023qxq}, our analysis  focuses on assessing how variations in these parameters impact the ability to satisfy the observed mass and radius constraints under the preferred Scenarios C and B. While results for Scenario A are included for completeness, they are clearly presented with the strong caveat that this scenario has been statistically ruled out and is therefore not a physically favored configuration.

This paper is organized as follows. In Sec.~\ref{sec:formalism} we give the detailed formalism of the equation of state. In Sec.~\ref{sec:results} we show the numerical results for different observables. Finally, we summarize in Sec.~\ref{sec:conclusion}.

\section{Formalism}\label{sec:formalism}

For the description of the hadronic sector in hybrid stars (HSs), we adopt the relativistic mean field (RMF) model, the specifics of which are provided in \cite{Glendenning:1997wn}. 
\begin{equation}
\begin{aligned}
    &\mathcal{L}=\bar{\psi}(i\gamma_{\mu}\partial^{\mu}-m_N)\psi+\frac{1}{2}(\partial_{\mu}\sigma\partial^{\mu}\sigma-m_{\sigma}^2\sigma^2)-\frac{1}{4}\omega_{\mu\nu}\omega^{\mu\nu} +\frac{1}{2}m_{\omega}^2\omega_{\mu}\omega^{\mu}-\frac{1}{4}\rho_{\mu\nu}\rho^{\mu\nu}+\frac{1}{2}m_{\rho}^2\vec{\rho_{\mu}}\vec{\rho^{\mu}} +\sum_{l= e,\mu}\bar{\psi_l}(i\gamma_{\mu}\partial^{\mu}-m_l)\psi_l \\ &+ (g_{\sigma}\bar{\psi}\sigma\psi -g_{\omega}\bar{\psi}\gamma_{\mu}\omega^{\mu}\psi -\frac{1}{2} g_{\rho}\bar{\psi}\gamma_{\mu}\vec{\tau}.\vec{\rho^{\mu}}\psi)-\frac{1}{3}bm(g_{\sigma}\sigma)^3 -\frac{c}{4} (g_{\sigma}\sigma)^4  +\Lambda_{\omega \rho}g_{\omega}^2(\omega_{\mu}\omega^{\mu})(g_{\rho}^2\vec\rho_{\mu}\vec{\rho^{\mu}})+\frac{\zeta}{24}(g_\omega^2\omega_\mu\omega^\mu)^2 ~. \\ &+
g_\sigma g_\omega^2\sigma\omega_\mu\omega^\mu
\left(\alpha_1+\frac{1}{2}{\alpha_1}'g_\sigma\sigma\right)
+ g_\sigma g_\rho^2\sigma\vec{\rho}_\mu\vec{\rho}^\mu
\left(\alpha_2+\frac{1}{2}{\alpha_2}'g_\sigma\sigma\right) 
\end{aligned}
\end{equation} 
In this work we have considered six different types of
parameterizations.
\begin{itemize}
    \item Type 1 : $g_{i=\sigma,\omega,\rho}=\text{constant},~\Lambda_{\omega\rho}=0,~\zeta=0, \alpha_1=\alpha_2=\alpha_{1}^{'}=\alpha_{2}^{'}=0 $ (GM1-\cite{Glendenning:1991es})
    \item Type 2 : $g_{i=\sigma,\omega,\rho}=\text{constant},~\Lambda_{\omega\rho}=0,~\zeta\not=0,\alpha_1=\alpha_2=\alpha_{1}^{'}=\alpha_{2}^{'}=0$ (TM1\cite{Geng:2003pk})
    \item Type 3 : $g_{i=\sigma,\omega,\rho}=\text{constant},~\Lambda_{\omega\rho}\not=0,~\zeta=0, \alpha_1=\alpha_2=\alpha_{1}^{'}=\alpha_{2}^{'}=0$ (NL3$\rho\omega$ \cite{Horowitz:2000xj})   
    \item Type 4 : $g_{i=\sigma,\omega,\rho}=\text{constant},~\Lambda_{\omega\rho}\not=0,~\zeta\not=0, \alpha_1=\alpha_2=\alpha_{1}^{'}=\alpha_{2}^{'}=0$ (FSU2R \cite{Tolos:2016hhl})
    \item Type 5 : $g_{i=\sigma,\omega,\rho}=\text{constant},~\Lambda_{\omega\rho}\not=0,~\zeta\not=0, \alpha_1=\alpha_2=\alpha_{1}^{'}=\alpha_{2}^{'}\not=0$ (BSR8 \cite{Dhiman:2007ck,Agrawal:2010wg})    \item Type 6 : $g_{i=\sigma,\omega,\rho}=\text{Density Dependent},~\Lambda_{\omega\rho}=0,~\zeta=0$ (DD2 \cite{Typel:2009sy})
\end{itemize}

\begin{table}[!ht]
\caption{The nuclear matter properties at saturation density $\rho_{_{0}}$ for different hadronic models}
\setlength{\tabcolsep}{20.0pt}
\begin{tabular}{ccccccc}
\hline
\hline
Model & $\rho_{0}$ & $B/A$ & $K_{sat}$ &$\text{m}^*/\text{m}$  & $\text{E}_{\text{sym}}$ &$\text{L}_{\text{sym}}$\\
& $({\rm fm}^{-3})$ & (MeV) & (MeV) &  &(MeV) & (MeV)\\ \hline
GM1 & 0.153 & $-$16.30 & 300.0 & 0.70 & 32.5 & 93.9 \\
TM1 & 0.145 & $-$16.26 & 281.2 & 0.63& 36.89 & 110.79\\
NL3$\rho \omega$4 & 0.148 & $-$16.40 & 231.2 & 0.59 & 31.10 &68.2\\
FSU2R & 0.150 & $-$16.28 & 238.0 & 0.593 & 30.2 &44.30\\
BSR8 & 0.147 & $-$16.04 & 230.95 & 0.61 & 31.08  & 60.25 \\
DD2 & 0.149 & $-$16.02 & 242.72 & 0.56 & 31.67 & 55.04 \\
\hline
\hline
\end{tabular}
\label{tab:1}
\end{table}

This classification summarizes various types of relativistic mean-field (RMF) models based on the functional form of the meson-nucleon coupling constants and the presence of nonlinear interaction terms. In \textbf{Type 1}, the meson couplings $g_{\sigma}, g_{\omega}, g_{\rho}$ are constant, and no nonlinear self-couplings or mixed interactions are included (e.g., GM1~\cite{Glendenning:1991es}). \textbf{Type 2} retains constant couplings but includes a non-zero $\zeta$ parameter, which represents the self-interaction of the $\omega$ meson (e.g., TM1~\cite{Geng:2003pk}). \textbf{Type 3} introduces a mixed isoscalar-isovector coupling through a non-zero $\Lambda_{\omega\rho}$, while setting $\zeta = 0$ (e.g.,~\cite{Horowitz:2000xj}). \textbf{Type 4} includes both $\Lambda_{\omega\rho}$ and $\zeta$ terms (e.g., FSU2R~\cite{Tolos:2016hhl}). In \textbf{Type 5}, additional nonlinear cross-coupling terms, denoted by the $\alpha$-parameters ($\alpha_1, \alpha_2, \alpha_1', \alpha_2'$), are also introduced to increase the flexibility of the model (e.g., BSR8~\cite{Dhiman:2007ck,Agrawal:2010wg}). Finally, \textbf{Type 6} represents models where the meson-nucleon couplings are explicitly density-dependent, as in the DD2 parametrization~\cite{Typel:2009sy}, with no self or mixed couplings ($\Lambda_{\omega\rho} = \zeta = 0$). This systematic classification aids in comparing the role of different interaction terms in shaping the nuclear matter equation of state.
Models like NL3$\rho\omega4$, TM1, and GM1 are included as benchmark cases for stiff hadronic equations of state, allowing us to explore extreme limits of hybrid star properties. Additionally, their inclusion helps illustrate the classification of RMF models based on meson interaction structures, as described earlier.

The quark sector is modeled using the simplified Constant Speed of Sound (CSS) parametrization, which characterizes the hadron-quark phase transition and the quark core. The CSS model is a phenomenological idealization, wherein the speed of sound $ C_s $ is treated as constant. However, in realistic systems, $C_s^2 $ generally depends on the density.
While this approximation allows for analytical simplicity, it may not accurately capture 
 the behavior in regimes where $ C_s^2 $ varies significantly with density. This model is defined by three parameters: the transition density, $\rho_{tr}$; the energy density discontinuity, $\Delta\varepsilon$; and the squared speed of sound, $c_s^2$. We have adopted the parametrization introduced by \cite{Pal:2025skz,PhysRevD.88.083013,Tsaloukidis:2022rus,Laskos-Patkos:2023cts}. Below the transition pressure($P_{tr}$) or transition density ($\rho_{tr}$), we consider hadronic EoS($\varepsilon_H$)  whereas above the transition point, we use the constant speed of sound parametrizations for the quark matter. The energy density below and above the transition density is as follows:

\begin{equation}
    \varepsilon(P)=\begin{cases}
    \varepsilon_H, & \text{if } P \leq P_{tr} \\
    \varepsilon(P_{tr}) + \Delta \varepsilon + \frac{1}{C_s^2}(P-P_{tr}), & \text{if } P > P_{tr}
    \end{cases}
\end{equation}

\section{Results}
\label{sec:results}

\begin{figure*}[htp] 
    \centering
   \includegraphics[width=0.45\textwidth]{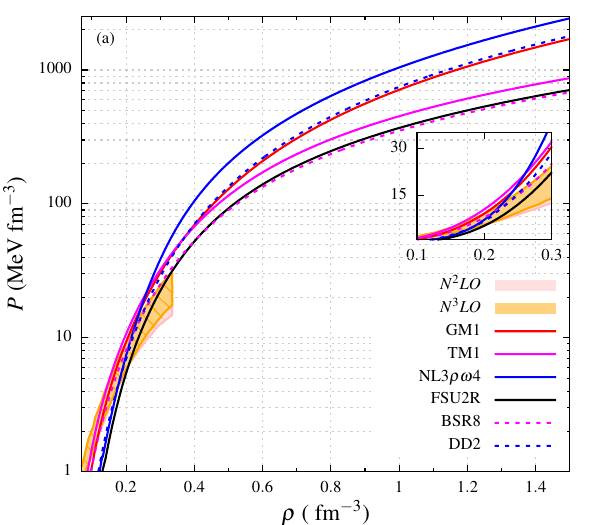}
   \includegraphics[width=0.45\textwidth]{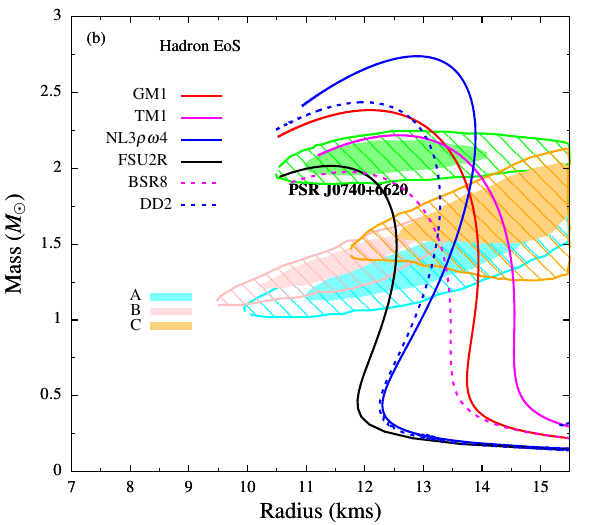}
    \caption{ Panel (a) displays the pressure–baryon density relations for various hadronic models, and panel (b) shows the corresponding mass–radius relationships }   
  \label{fig:hadron_eos_mr}
\end{figure*}

\begin{figure*}[htp] 
    \centering
     \includegraphics[width=0.45\textwidth]{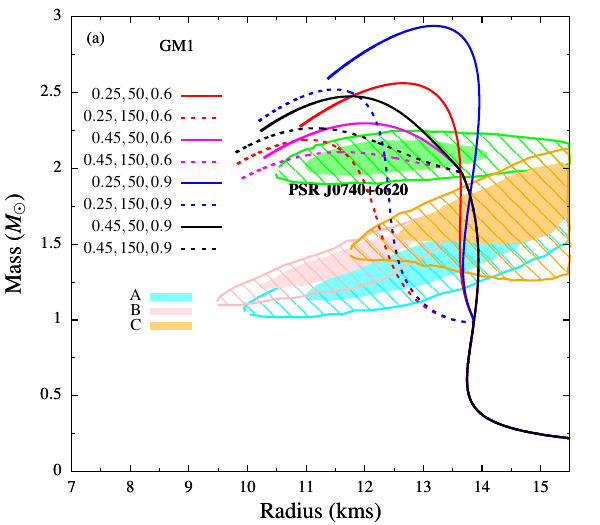}
          \includegraphics[width=0.45\textwidth]{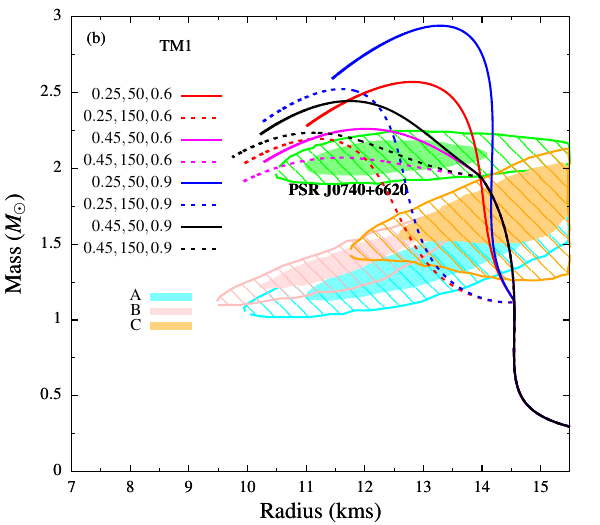}
          \includegraphics[width=0.45\textwidth]{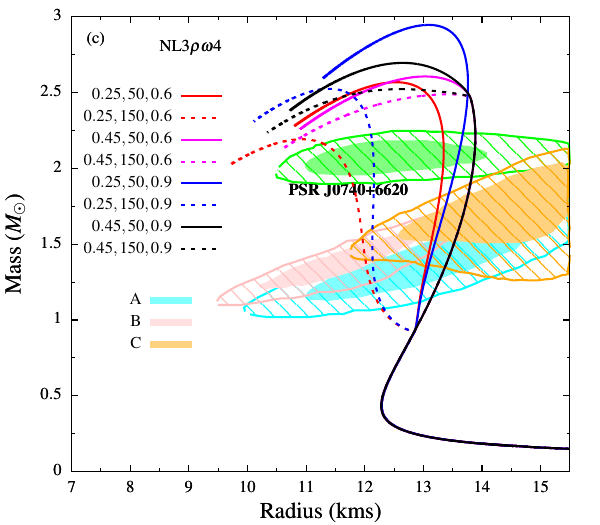}
     \includegraphics[width=0.45\textwidth]{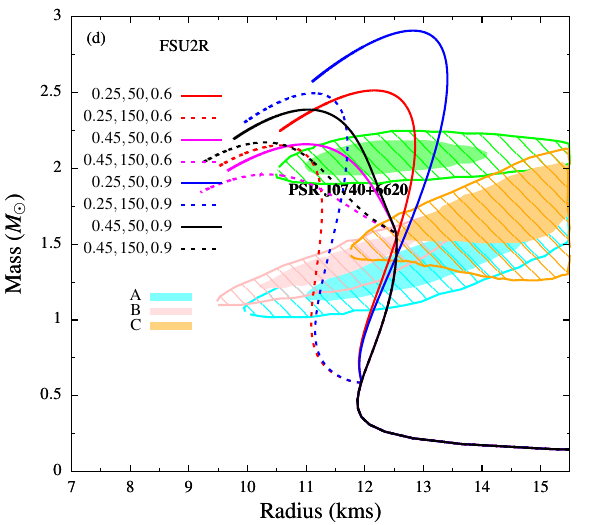}
               \includegraphics[width=0.45\textwidth]{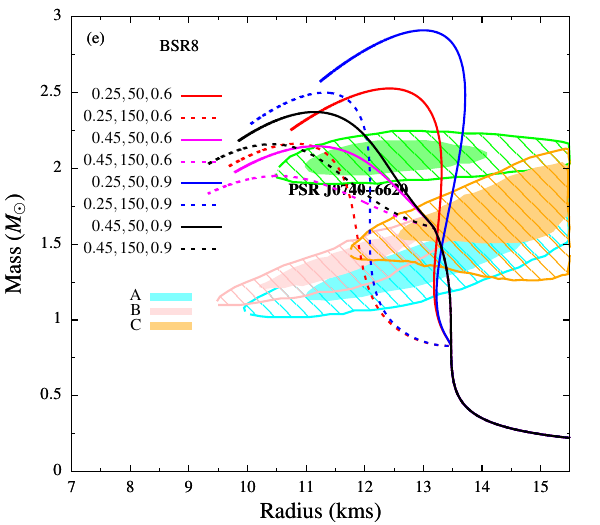}
     \includegraphics[width=0.45\textwidth]{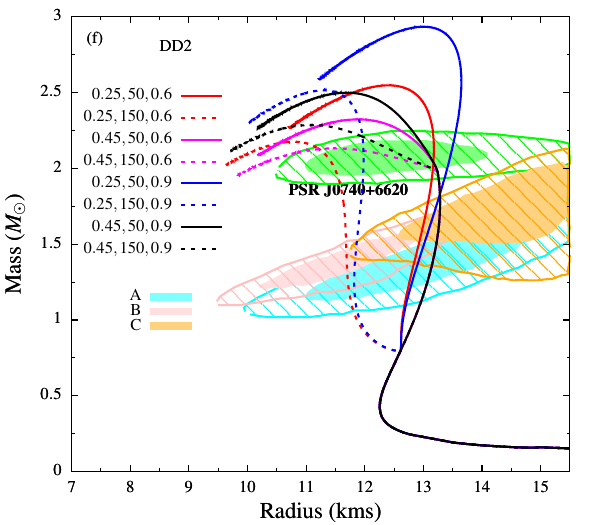}

    \caption{Mass–radius relations for hybrid star configurations, based on the hadronic EoS shown in Fig.~\ref{fig:hadron_eos_mr}, are obtained using the CSS parameterization.}   
    \label{fig:mr_hs_six}
\end{figure*}

The initial part of the this section, which presents  Fig.\ref{fig:hadron_eos_mr} and Fig.\ref{fig:mr_hs_six}, plays an essential role in establishing the context and framework for the subsequent analysis.
In Fig.\ref{fig:hadron_eos_mr}(a), we have plotted the six hadron EoS  which are constructed based on RMF theory. The EoS are in accordance with the results obtained upto next-to-next-to-next-to-leading order
($N^{3} LO$) and next-to-next-to-leading order ($N^{ 2} LO$)
in the chiral effective field theory ($\chi$ EFT) expansion \cite{Drischler:2020fvz}. The properties of nuclear matter at saturation density are presented in Table~\ref{tab:1}. The figure displays the relative stiffness of the EoS and shows that three EOS (GM1, NL3$\rho \omega$4, DD2) are relatively more stiffer than the other three (TM1, FSU2R, BSR8).  It is observed that BSR8 and FSU2R have slopes very close to each other. We  distinguish between those models based on their consistency with modern nuclear constraints (e.g., $K_0$, $L_0$, and $\chi$EFT). Specifically, although models such as NL3$\rho\omega$4, GM1, and TM1 are in known tension with these constraints, we retain them as representative examples of extremely stiff EoSs. Their inclusion serves to highlight how such stiffness influences the onset of phase transitions and the upper bounds on neutron star masses. In Fig.\ref{fig:hadron_eos_mr}(b), we plot the mass-radius results for these six EoS and try to examine the compatibility with the observational constraints of the  three different categories of the new NICER measurements of J003 0+0451  as well as with the pulsar PSR J0740+6620. It is observed that all the six EoS satisfy the constraints from the category A, B and C of PSR J003 0+0451  except the EoS based on TM1 parametrization which  does not satisfy the constraints imposed by the category B of the new NICER measurement of PSR J0030+0451. These EoS also satisfy the constraint from PSR J0740+6620 ; the EoS from NL3, TM1, NL3$\rho\omega$, DD2 and GM1 prametrization have maximum mass  greater than those from FSU2R and BSR8 which barely satisfy the 2 $M_{\odot}$ constraint. The EoS based on TM1 have maximum mass value just above 2 $M_{\odot}$.

In Fig.~\ref{fig:mr_hs_six}, we plot the M-R diagram for the six different hadron EoS after phase transition with the CSS parameters as described in the formalism section. 
In our study, we investigate two typical transition density values: $0.25$ and \(0.45~\text{fm}^{-3}\), along with energy density jumps of \(50\) and \(150~\text{MeV fm}^{-3}\). For the quark phase, we consider squared speed of sound values, \(C_s^2\) = \(0.6\) and \(0.9\) for each energy density gap.

Results from stiffer hadronic models such as GM1, TM1 and $NL3\rho\omega4$ are also presented in Fig.~\ref{fig:mr_hs_six}(a,b and c) for completeness. Their inclusion primarily serves to highlight the model dependence and the effect of hadronic stiffness on hybrid star properties. 
 In Fig.~\ref{fig:mr_hs_six}(a),  we have considered the hybrid star results corresponding to the  GM1 parametrization for the hadron sector.  For $\rho_{\text{tr}}$ =0.25 $\text{fm}^{-3}$, all the four M-R plots (for two energy density gaps and two $C_s^2$ ) satisfy the observational constraints of category A and B of J003 0+0451 as well as the constraint from PSR J0740+6620. For the higher transition density  $\rho_{\text{tr}}$ = 0.45 $\text{fm}^{-3}$, the results for hybrid branch lie well  above the constraints of mass from the new NICER measurements  of PRS  J0030+0451 in all the categories A, B and C. They however satisfy the constraints of mass and radius of the pulsar PSR J0740+6620. In Fig.~\ref{fig:mr_hs_six}(b) we have plotted the results corresponding to the hadron EoS based on the TM1 parametrization. It is observed that the results for mass and radius and their compatibility with the different observational constraints for TM1 resemble closely  that of GM1 (Fig.~\ref{fig:mr_hs_six}(a)) parametrization. The results from the NL3 parametrization as displayed in Fig.~\ref{fig:mr_hs_six}(c) also resemble the previous two to a good extent, one striking difference being that the transition mass at $\rho_{\text{tr}}$  = 0.45 $\text{fm}^{-3}$ is roughly 2.5 $M_{\odot}$ in contrast to 2 $M_{\odot}$ in the previous two cases.

Among the hadronic models considered, DD2, FSU2R, and BSR8 are consistent with known nuclear matter constraints thus provide a more realistic foundation for hybrid star modeling. 
{\color{black}In case of FSU2R parametrization in Fig.~\ref{fig:mr_hs_six}(d), the transition mass at $\rho_{\text{tr}}$  = 0.45 $\text{fm}^{-3}$ is much lower and roughly equal to 1.5 solar mass and the transition zone overlaps with the category B of the new NICER measurements. The hybrid star formed at the higher transition density fulfills the constraints of category B and C; the one formed at the lower transition density fails to meet the constraints of category C of PSR J0030+0451.  The results of the BSR8 parametrization resemble closely that of FSU2R and this is evident from Fig1(a) where the EoS are found to be close to each other. The transition mass of the DD2 parametrization for $\rho_{\text{tr}}$ = 0.45 $\text{fm}^{-3}$ is about 2 solar mass and is close to that of GM1 and TM1 parametrization. The results for $\rho_{\text{tr}}$ = 0.25 $\text{fm}^{-3}$ for the higher energy density gap (150 $\text{MeV~fm}^{-3}$) does not fulfill the criteria of category C of PSR J0030+0451. 
The maximum mass  for all the six hadron EoS  for the lower transition density (0.25) and for the lower energy gap $\Delta\varepsilon$=50 $\text{MeV~fm}^{-3}$ for $C_s^2$ =0.9 is close to 3 solar mass. The observational constraints of mass and radius from  PSR J0740+6620 is satisfied by almost all the hybrid EoS for the entire range of the CSS parameters used.}

From a physical perspective, our results show clear trends of the hybrid star properties being affected by the parameters of the CSS quark matter model. For a fixed transition density $\rho_{\text{tr}}$, increasing the energy density discontinuity $\Delta\varepsilon$ tends to suppress the appearance of quark matter, resulting in a decrease in the maximum mass of the hybrid star. Conversely, smaller gaps allow a smoother transition and support more massive hybrid configurations. Similarly, increasing the speed of sound squared in the quark phase ($C_s^2$) enhances the stiffness of the quark matter EoS, which in turn increases the maximum mass. Additionally, lowering the transition density allows quark matter to appear at lower central densities, enabling even canonical-mass stars to access the hybrid branch and pushing the maximum mass higher. 
Overall, our results show that the observational constraints from PSR J0740+6620 and NICER can be successfully satisfied by a wide range of CSS parameters, particularly for physically consistent hadronic EoS models, thereby providing support for the hybrid star hypothesis within a realistic nuclear physics framework.
It is important to highlight that in the current analysis, all three NICER scenarios (A, B, and C) are weighted equally, even though the original NICER+XMM study by \cite{Vinciguerra:2023qxq} has a strong statistical preference. Their results indicate that Scenario C (`PDT-U') is strongly favored, Scenario B (`ST+PDT') is a less probable alternative, and Scenario A (`ST+PST') is statistically excluded. Consequently, Scenario C should be viewed as the most physically relevant, while the implications of Scenario A must be interpreted with caution, as they rely on a surface model that is inconsistent with existing observational constraints.

\subsection*{Constraints on CSS Parameters from Recent NICER Mass-Radius Measurements }

In this section, we investigate the range of CSS parameter values that can simultaneously account for the observed masses and radii across Scenarios A, B, and C. The details of these scenarios are as follows:
Scenario A: $M = 1.37^{+0.17}{-0.17},M{\odot}$, $R = 13.11^{+1.30}{-1.30}$~km;
Scenario B: $M = 1.20^{+0.14}{-0.11},M_{\odot}$, $R = 11.16^{+0.90}{-0.80}$~km;
Scenario C: $M = 1.41^{+0.20}{-0.19},M_{\odot}$, $R = 13.12^{+1.35}_{-1.21}$~km.
These are considered alongside the observational constraints from PSR J0740+6620 ( $M=2.07\pm{0.07} M_{\odot}$, $R=12.49_{-0.88}^{+1.28}$ km).
The 1$\sigma$ measurements for Scenarios A and C are closely aligned in both mass and radius, whereas Scenario B displays notably different values, particularly in the radius. This difference plays a significant role in constraining the allowed region of CSS parameter space.
For this analysis, we adopt three representative values of the speed of sound squared in the quark phase, $C_s^2 = 0.6$, $0.75$, and $0.9$. The transition density $\rho_{\text{tr}}$ is varied in the range $0.25$–$0.45~\text{fm}^{-3}$, and the energy density discontinuity $\Delta\varepsilon$ is considered within $50$–$300~\text{MeV~fm}^{-3}$, with the upper limit depending on the specific hadronic EOS employed.
  \begin{figure*}[htp] 
    \centering
    \includegraphics[width=1.0\textwidth]{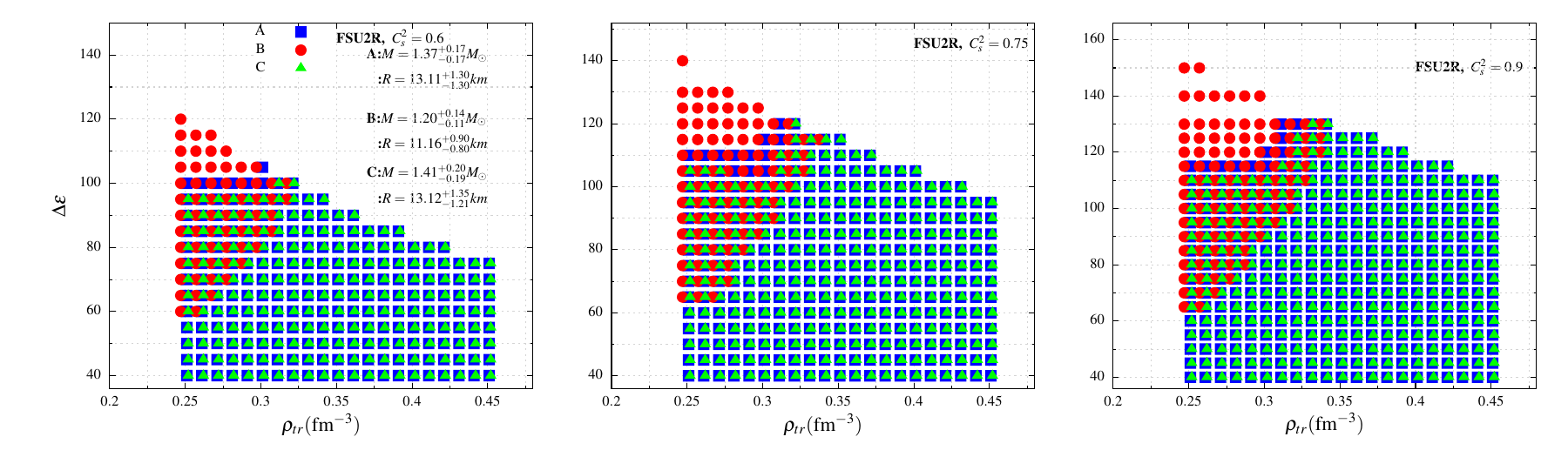}
    \includegraphics[width=1.0\textwidth]{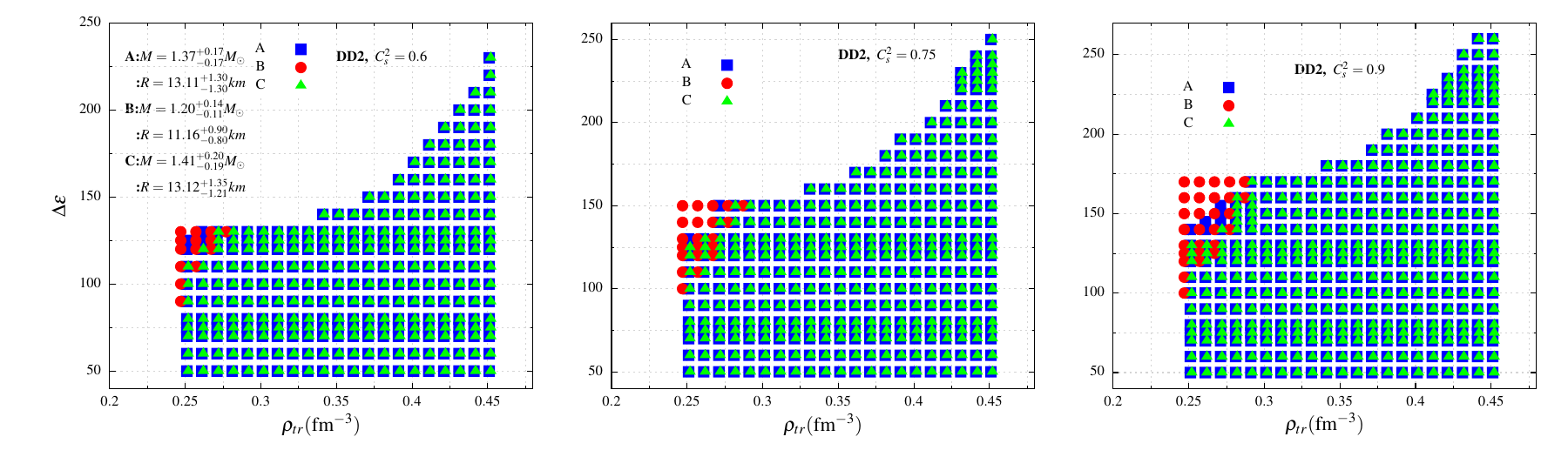}
  \caption{Transition density–energy density parameter space constrained by NICER measurements. The upper panel corresponds to the FSU2R EoS, while the lower panel corresponds to the DD2 EoS. The left and right panels represent $C_s^2 = 0.6$ and $C_s^2 = 0.9$, respectively. The details of the NICER constraints :
        \textbf{A}: $M=1.37_{-0.17}^{+0.17}M_{\odot}$,
        $R=13.11_{-1.30}^{+1.30}$ km;
        \textbf{B}: $M=1.20_{-0.11}^{+0.14}M_{\odot}$,
        $R=11.16_{-0.80}^{+0.90}$ km;
        \textbf{C}: $M=1.41_{-0.19}^{+0.20}M_{\odot}$,
        $R=13.12_{-1.21}^{+1.35}$ km. and along with PSR J0740+6620 : $M=2.07\pm{0.07} M_{\odot}$,
        $R=12.49_{-0.88}^{+1.28}$ km.
    }
    \label{fig:para_ABC}
\end{figure*}  
  In Fig.~\ref{fig:para_ABC}, we illustrate the viable parameter space of our CSS model, constrained by the updated NICER measurements of neutron star mass and radius as well as the constraint from the pulsar PSR J0740+6620.
  We have distinguished the different categories using distinct symbols: Category A is represented by a blue square, Category B by a red circle, and Category C by a green triangle.  For the hadronic EoS, we have used one density independent one, that is FSU2R and another is DD2 which is density dependent chosen for their physical consistency with nuclear saturation properties. Scenario A is included for completeness, but its important to note that it has been statistically ruled out and is not considered realistic.
  
  For the Eos based on the FSU2R parametrization,  results are quite close for the lower values of the energy density gap $\Delta \varepsilon$ for the entire range of $\rho_{\text{tr}}$ for three different values of $C_s^2$. Results are significantly different for the values of  $\Delta \varepsilon$ $>$ 100 $\text{MeV~fm}^{-3}$ in three different plots of Fig.~\ref{fig:para_ABC}(a). The effect of the transition density for the hadron to quark phase transition is reflected from these figures.~\ref{fig:para_ABC}.  For higher values of $\rho_{\text{tr}}$, its the category A and C which matches more with model calculations whereas category B agrees more at the lower density regime for higher values of the gap parameter. The constraint of 2$M_{\odot}$ from PSR J0740+6620 restricts the compatibility with higher energy gaps ($\Delta\varepsilon$ $>$ 150 $\text{MeV~fm}^{-3}$) for the entire range of $\rho_{\text{tr}}$ for all  three values of $C_s^2$ used in our calculations.  For the lowest energy density gap parameter, results from category A is reproduced as is observed from Fig.~\ref{fig:para_ABC}(a). For higher values of $C_s^2$ (0.75 and 0.9) results from our model satisfies more  the constraints from category A, B and C as compared to $C_s^2$  =0.6 at higher values of the energy density gap $\Delta \varepsilon$.

For the next hadronic EOS which is the density dependent DD2, our model results satisfy the criteria of Category B to a much lesser extent  as compared to FSU2R till $\Delta \varepsilon$ =100 MeV fm$^{-3}$.  Results for three different values of $C_s^2$ are quite close till the energy density gap of 130 MeV fm$^{-3}$ for the entire range of the transition densities. At higher energy gaps (140 and 150 MeV), results  differ as $C_s^2$ increases from 0.6 to 0.9 as is observed from lower panel of Fig.~\ref{fig:para_ABC}.  For higher  values of $C_s^2$ (0.75 and 0.9), higher values of the energy density gap $\Delta \varepsilon$ is allowed which satisfy the constraints. The results does not match the mass radius requirements of category  B in this transition density for energy density gap nearly greater than 150 MeV fm$^{-3}$ ($C_s^2=0.75$) and 175 MeV fm$^{-3}$ ($C_s^2=0.9$).

The distinct observational scenarios favor different regions of the parameter space due to their intrinsic sensitivity to the underlying hybrid star equation of state. This sensitivity arises from the way micro-physical properties, such as the characteristics of the hadron-quark phase transition and the interactions within quark matter determine the equation of state, which in turn governs macroscopic stellar features like mass and radius. Scenario B, which favors relatively smaller mass and radius, naturally selects lower transition densities and larger energy gaps compared to the other scenarios, since these features result in a softer equation of state that produces smaller stellar radii.
On the contrary, Scenario C, which permits relatively larger radii, is more consistent with delayed phase transitions and a stiffer equation of state, allowing the star to retain a more extended structure. Thus, each scenario constrains the equation of state in a physically meaningful way by linking observational trends to the underlying microphysics.

\subsection*{Constraints on CSS parameters from the maximum mass for  different hadronic models}

\begin{figure*}[htp] 
    \centering
    \includegraphics[width=0.30\textwidth]{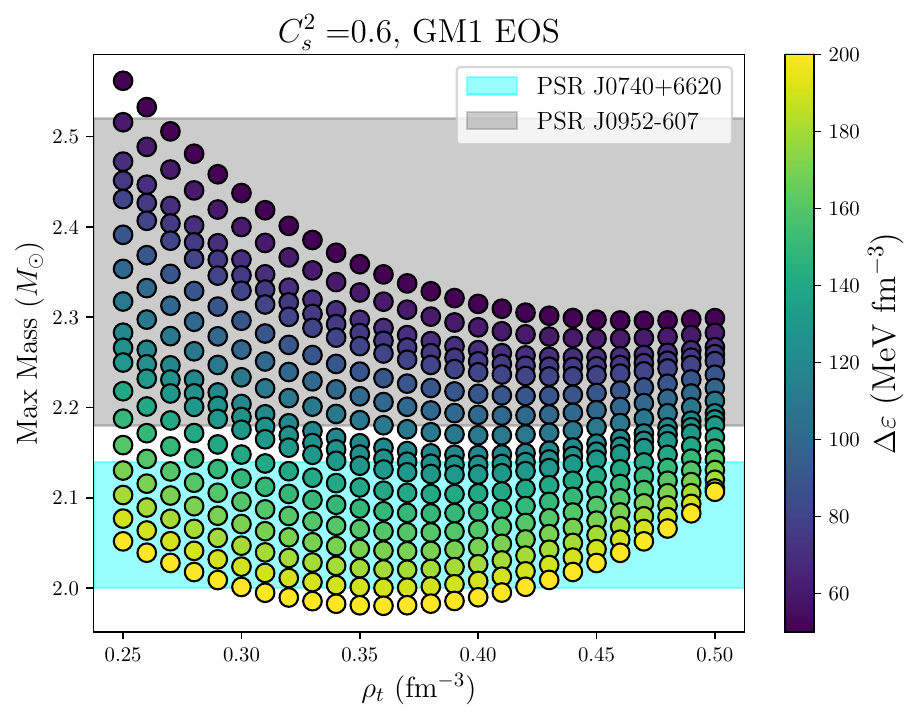}
    \includegraphics[width=0.30\textwidth]{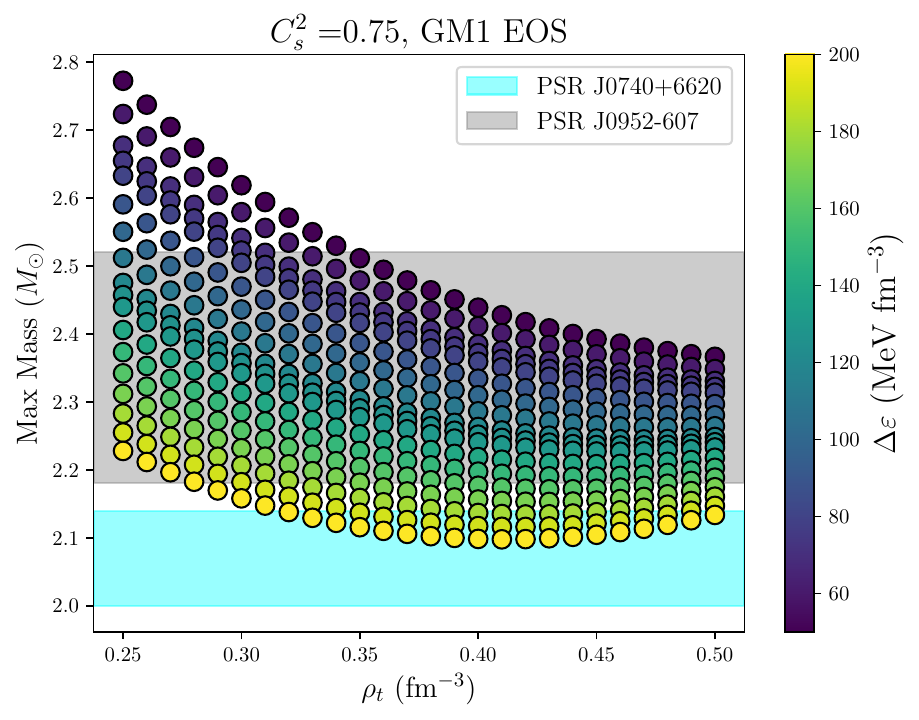}
     \includegraphics[width=0.30\textwidth]{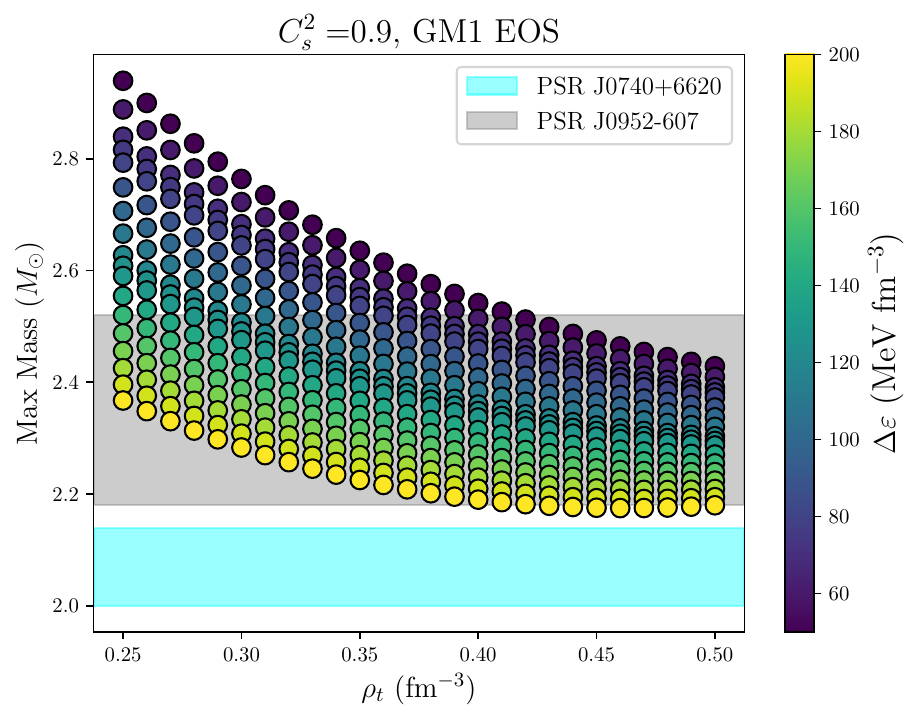}
      \includegraphics[width=0.30\textwidth]{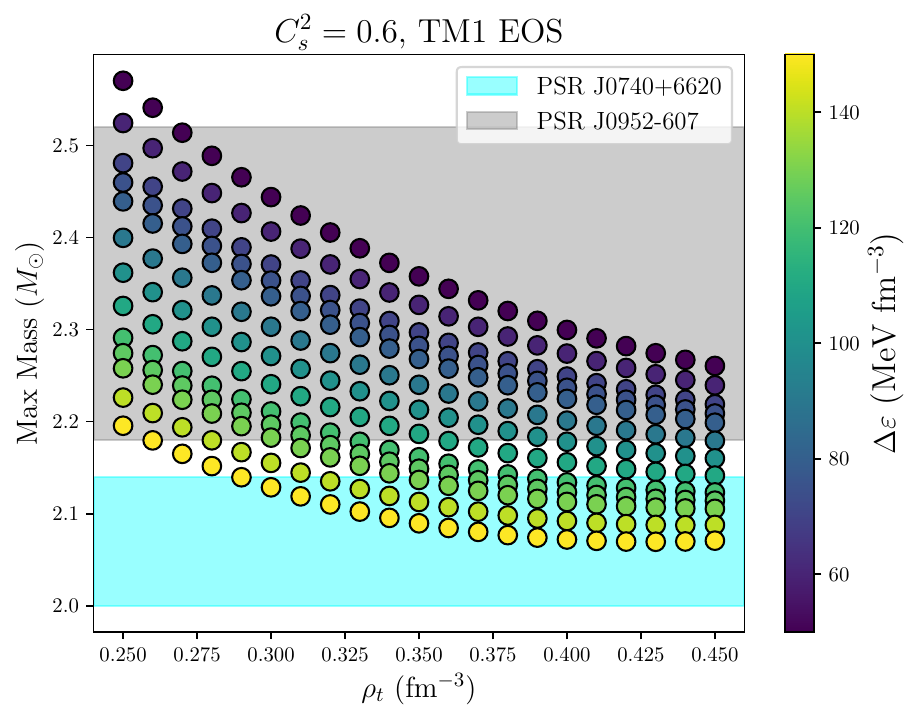}
    \includegraphics[width=0.30\textwidth]{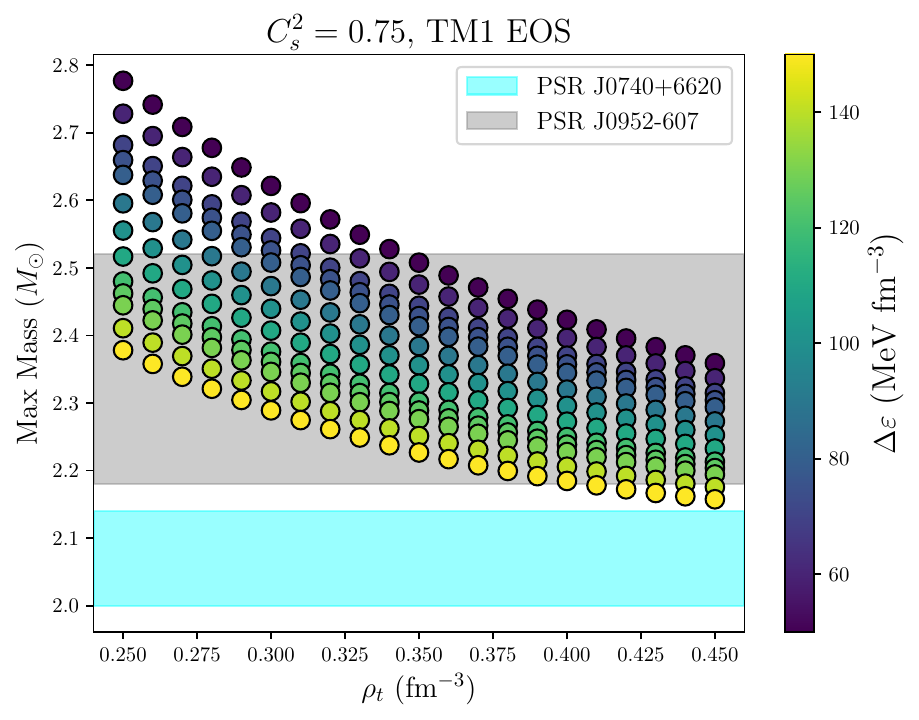}
     \includegraphics[width=0.30\textwidth]{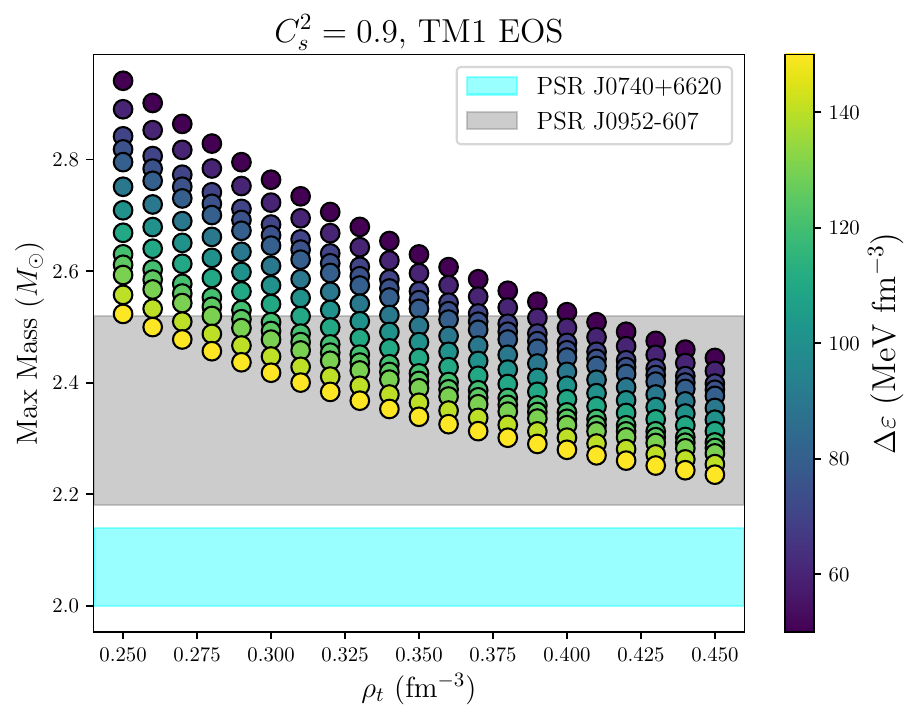}
     \includegraphics[width=0.30\textwidth]{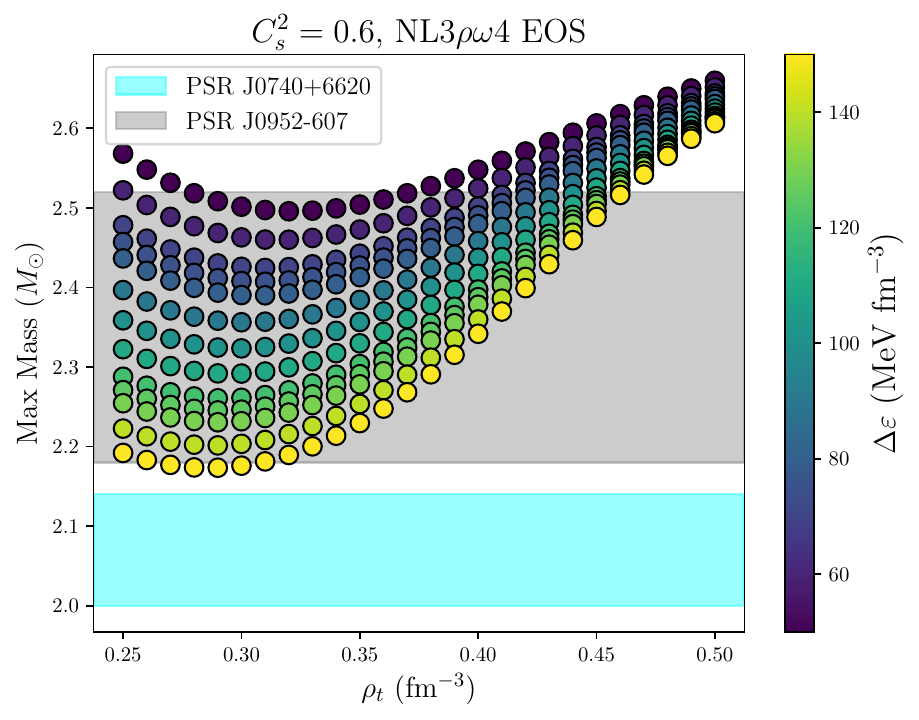}
    \includegraphics[width=0.30\textwidth]{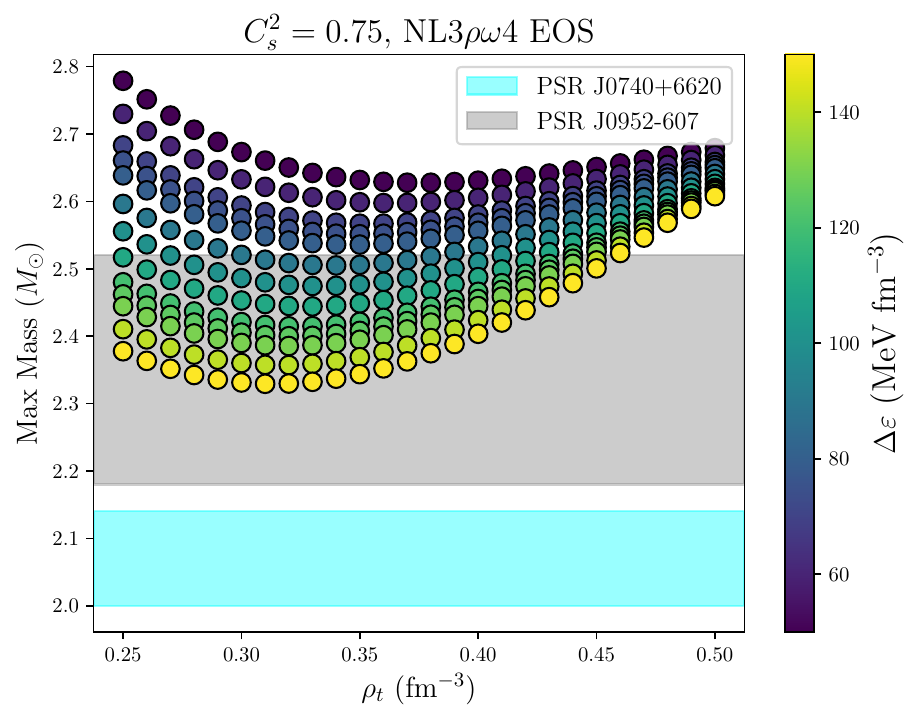}
     \includegraphics[width=0.30\textwidth]{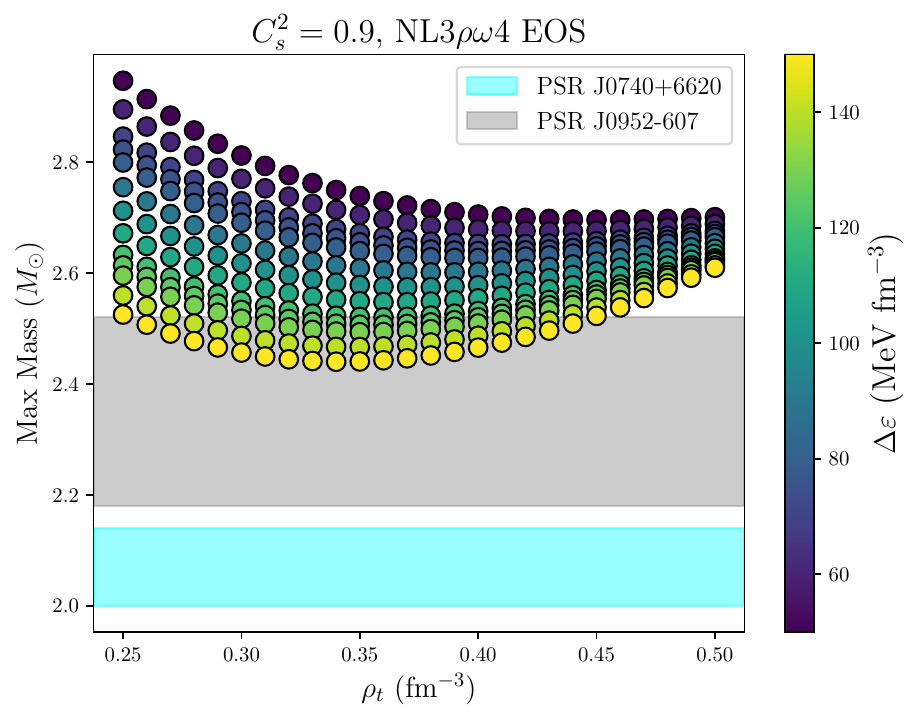}
     \includegraphics[width=0.30\textwidth]{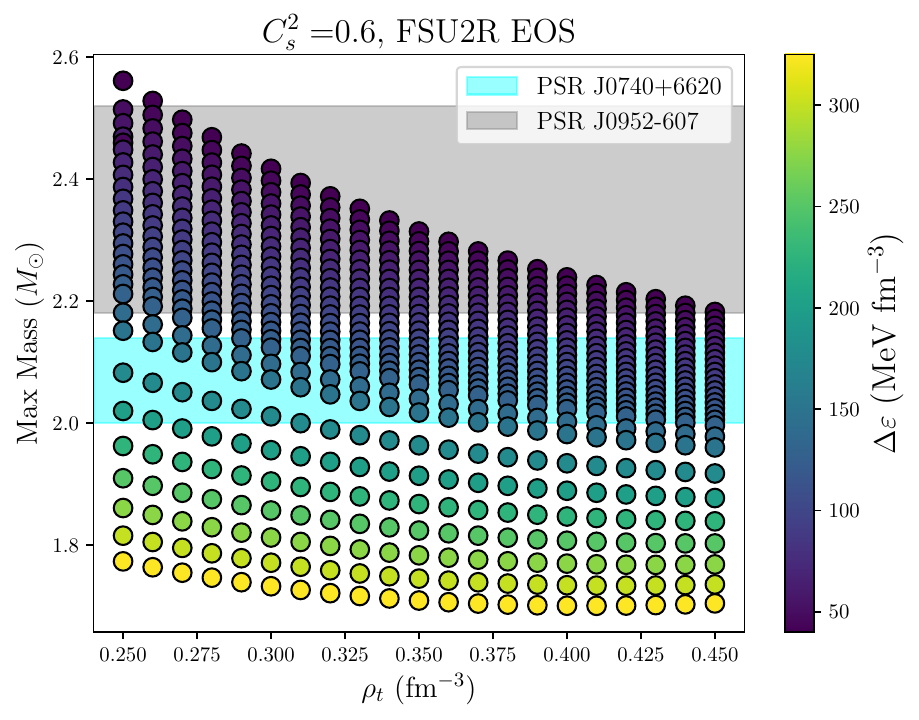}
    \includegraphics[width=0.30\textwidth]{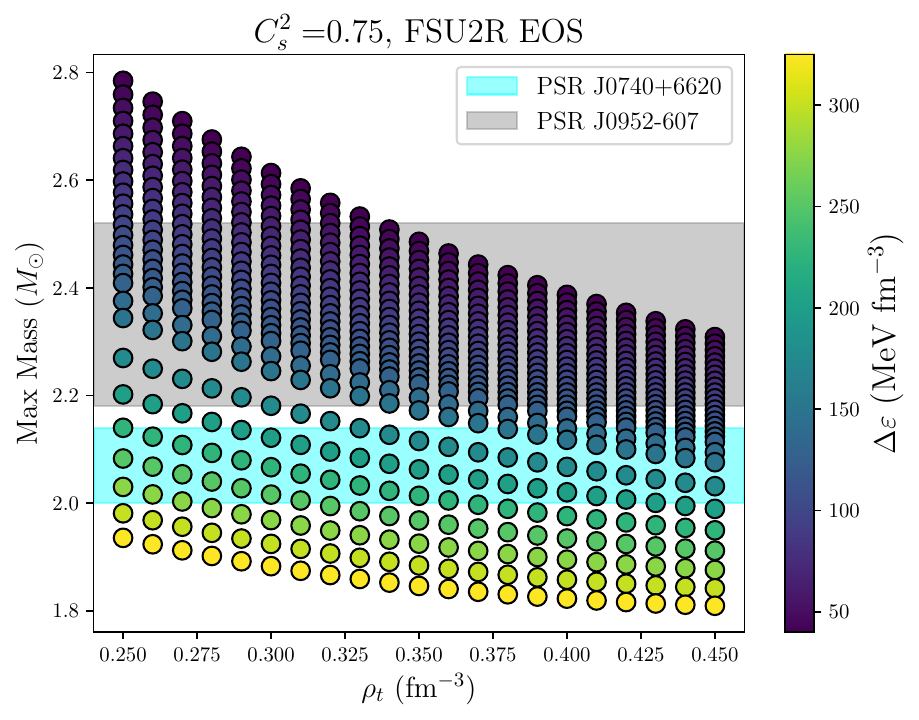}
     \includegraphics[width=0.30\textwidth]{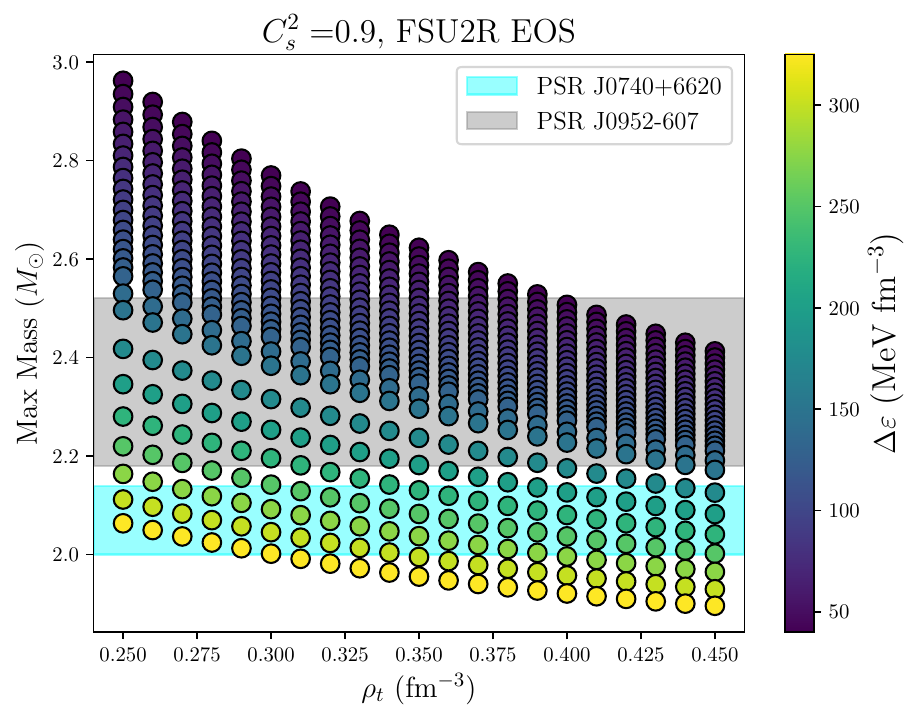}
          \includegraphics[width=0.30\textwidth]{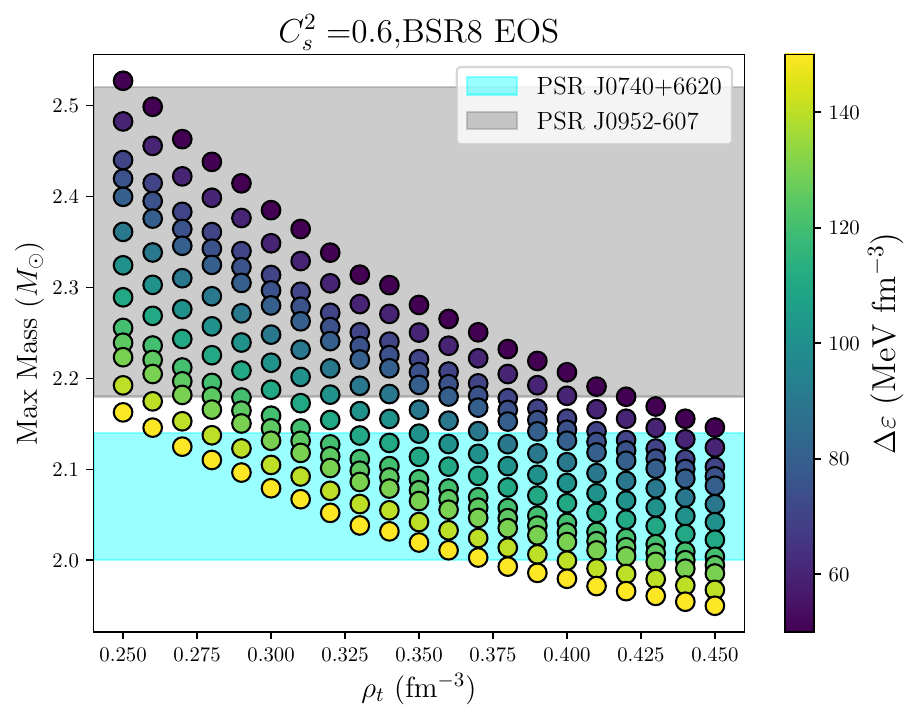}
     \includegraphics[width=0.30\textwidth]{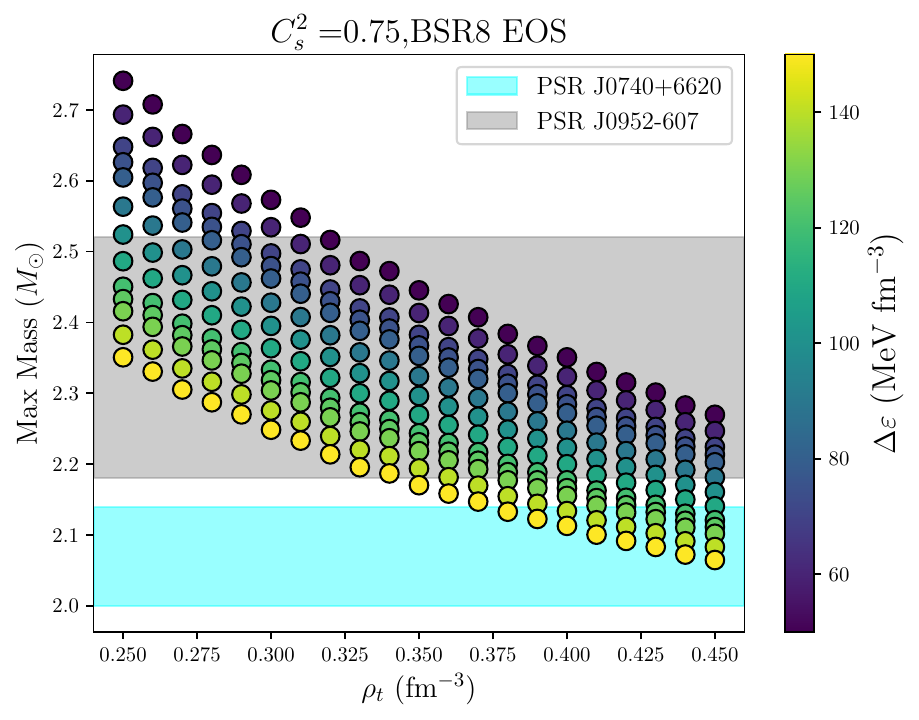}
     \includegraphics[width=0.30\textwidth]{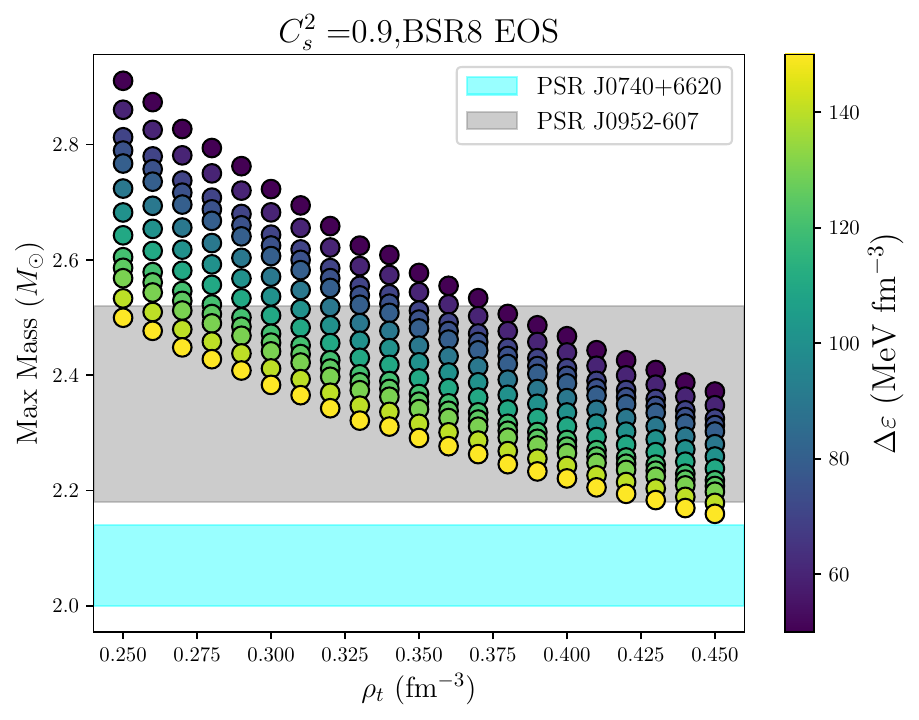}
    \caption{Maximum mass as a function of the transition density for different energy density gaps (indicated by the color bar) and various values of $C_s^2$: left panel ($C_s^2 = 0.4$), middle panel ($C_s^2 = 0.6$), and right panel ($C_s^2 = 0.9$) for the density independent hadronic EoS.
    }   
    \label{fig:mr_constraints}
\end{figure*}

\begin{figure*}[htp] 
    \centering
      \includegraphics[width=0.30\textwidth]{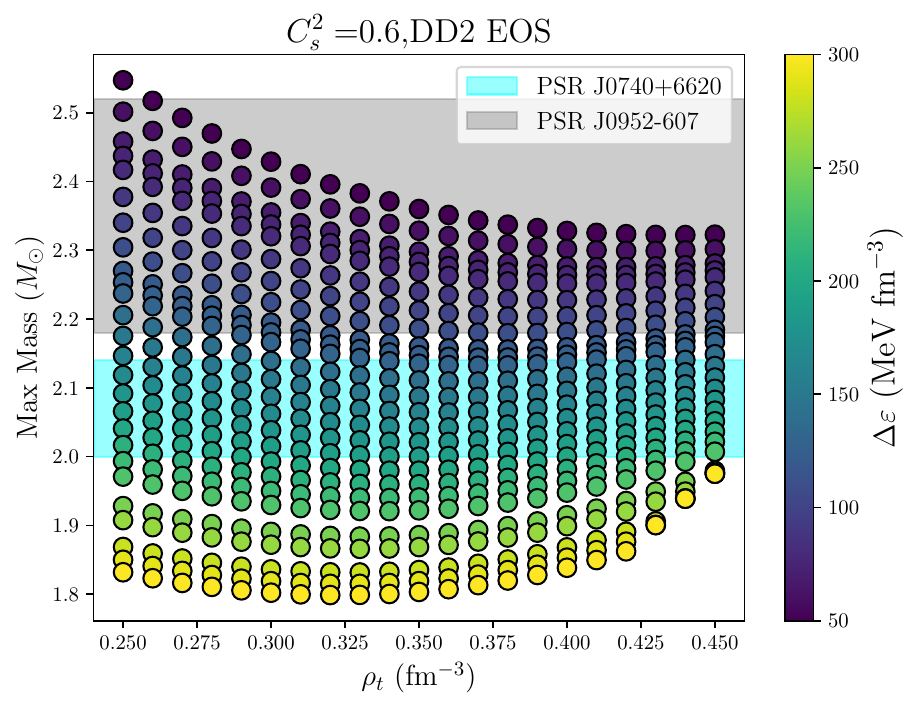}
     \includegraphics[width=0.30\textwidth]{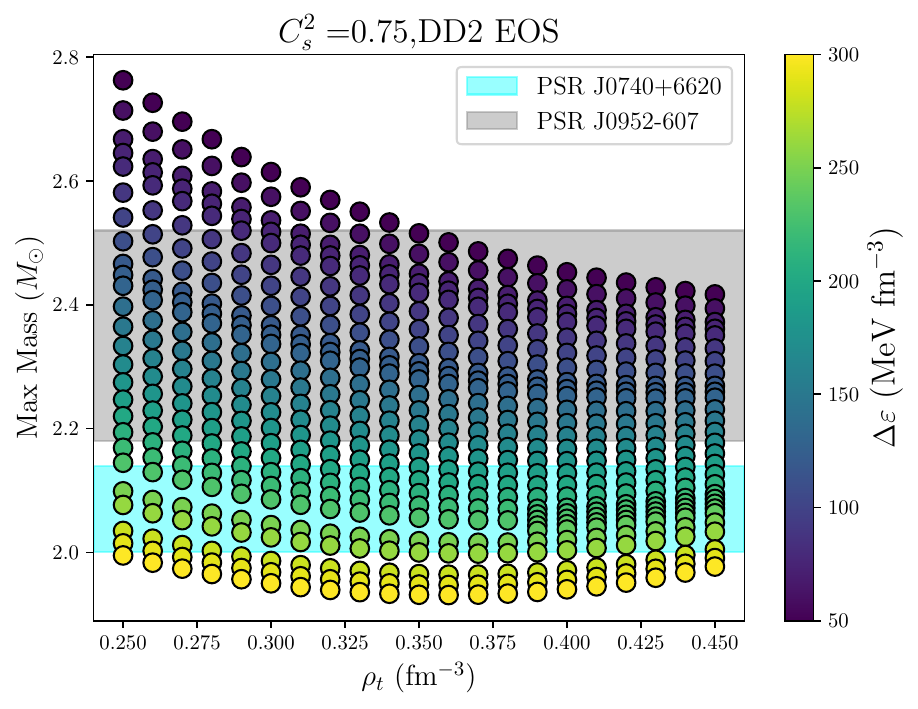}
     \includegraphics[width=0.30\textwidth]{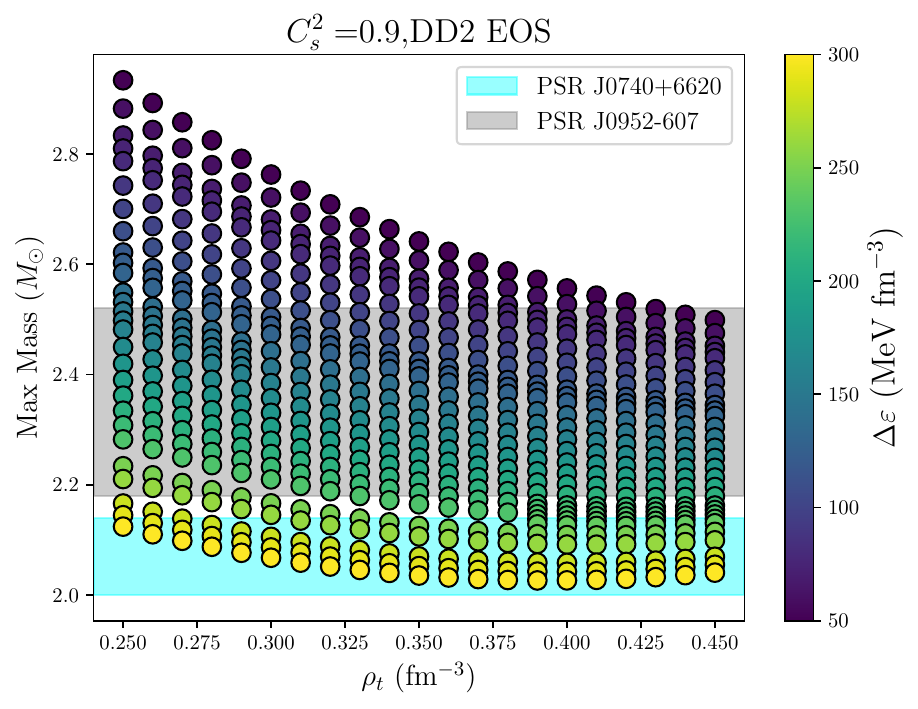}
    \caption{Maximum mass as a function of the transition density for different energy density gaps (indicated by the color bar) and various values of $C_s^2$: left panel ($C_s^2 = 0.4$), middle panel ($C_s^2 = 0.6$), and right panel ($C_s^2 = 0.9$) for the density dependent hadronic EoS.
    }   
    \label{fig:mr_DD_constraints}
\end{figure*}

\begin{figure*}[htp] 
    \centering
    \includegraphics[width=0.90\textwidth]{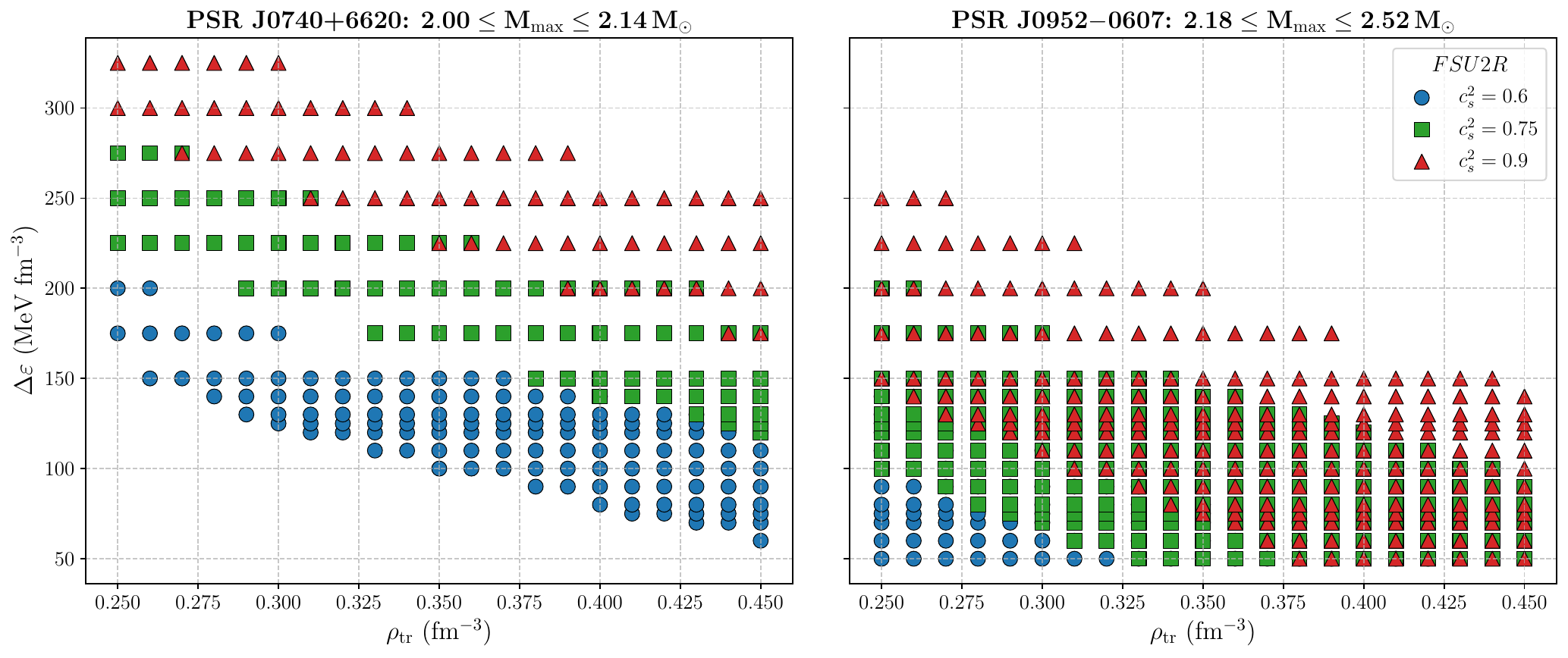}
        \includegraphics[width=0.90\textwidth]{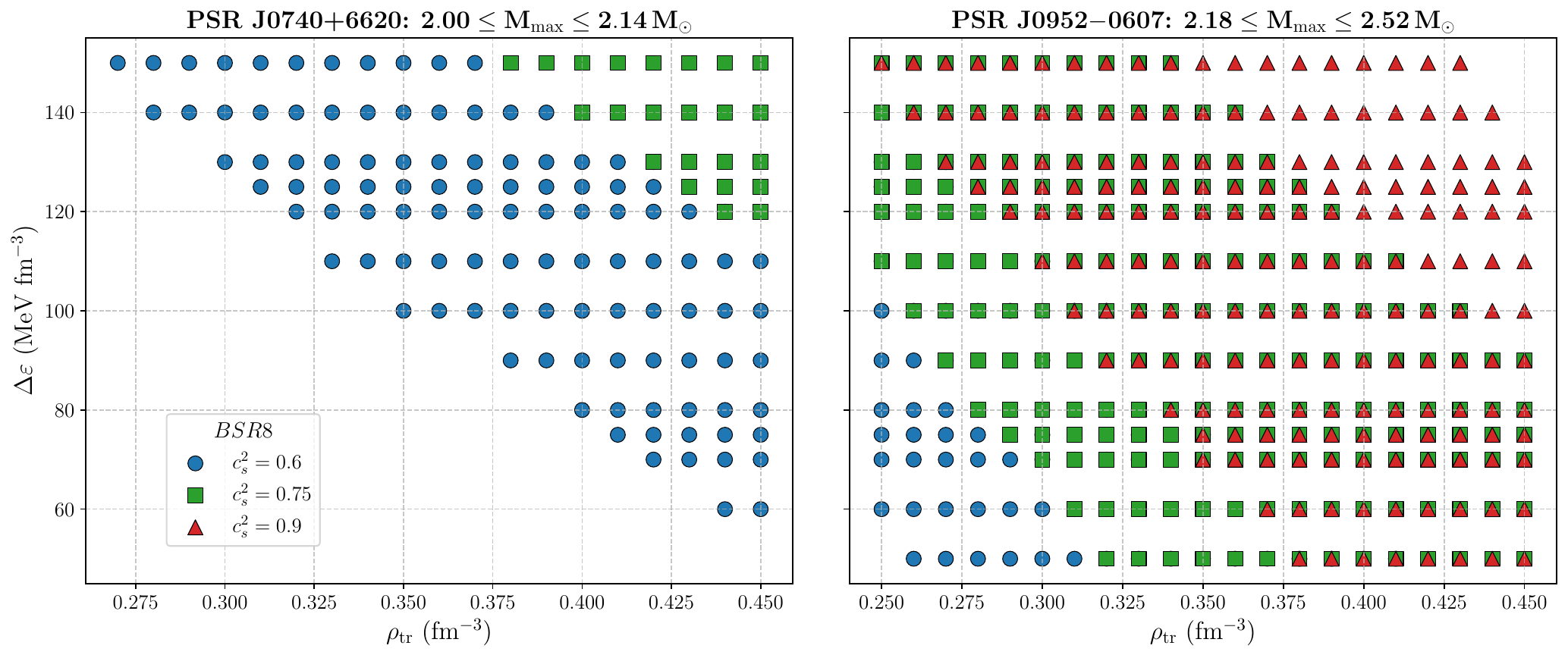}
              \includegraphics[width=0.90\textwidth]{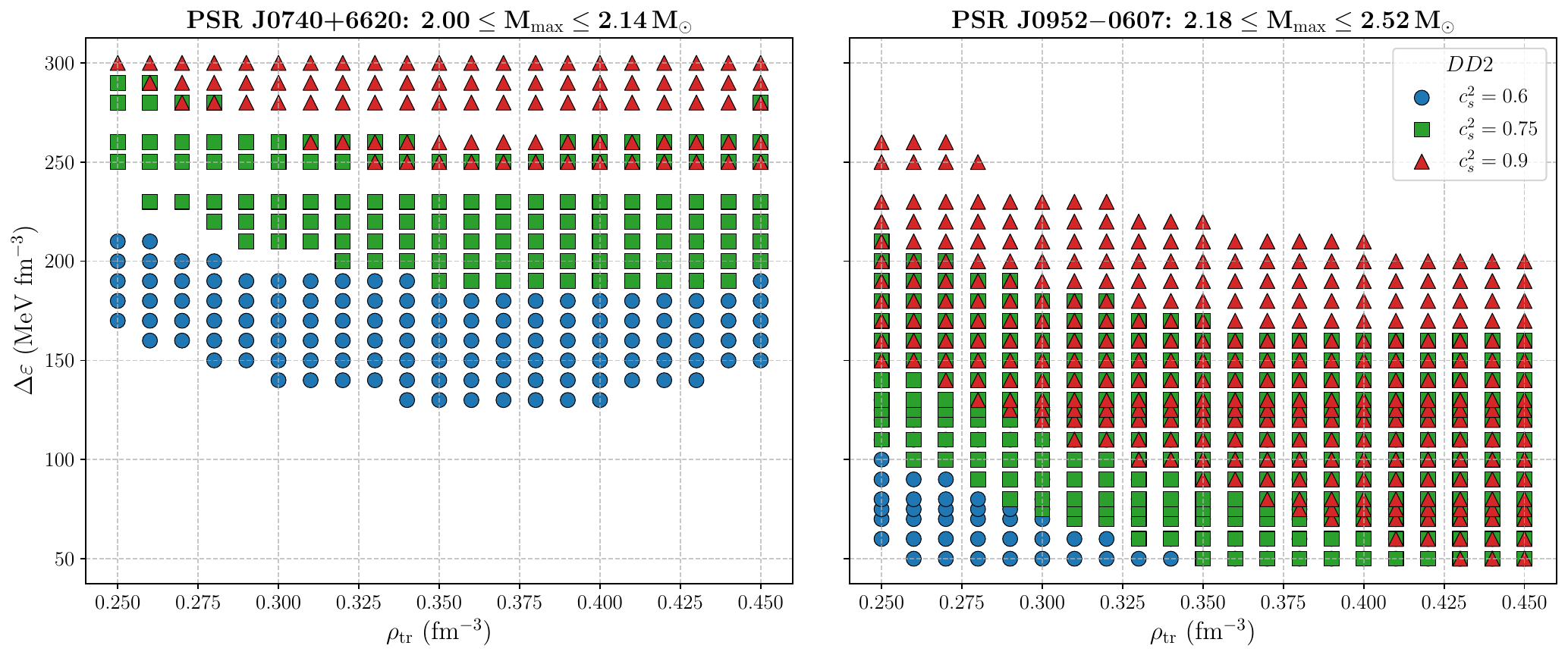}
    \caption{Allowed parameter space of the CSS  model from the constraints of PSR~J0740+6620 and PSR~J0952-0607 for three physically viable models. The $x$-axis represents the transition density $\rho_{\text{tr}}$, and the $y$-axis denotes the energy density discontinuity $\Delta \varepsilon$.}

    \label{fig:mr_psr_css_constraints}
\end{figure*}

We have also constrained the CSS parameters  ( energy density gap  $\Delta \varepsilon$ and $\rho_{\text{tr}}$ ) using the observational results from the pulsars PSR J0740+6620 and PSR J0952-607 by calculating the maximum mass obtained for these parameters. The calculations have been done for three values of $ C_s^2$;  0.6, 0.75 and  0.9 for each of the six hadronic EoS as is seen from Fig.~\ref{fig:mr_constraints}. For the EoS based on GM1 parametrization, results from the model calculations best reproduce the data for the lowest value of the speed of sound(0.6) as is observed from the Fig.~\ref{fig:mr_constraints}(a). As $C_s^2$ is increased to 0.9, the maximum mass obtained exceeds that from the pulsar PSR J0740+6620 for the entire range of density and energy gap used. At low transition densities and for the lower energy gaps, the maximum mass as obtained from the model exceeds the mass from PSR 0952-607 as well. The trend of the results from the EoS based on TM1 parametrization is similar to that of GM1 with the maximum mass being slightly higher in magnitude in case of TM1. 
  An interesting observation for the NL3$\rho\omega$ EoS is that for each  energy density  gap $\Delta \varepsilon$, the maximum mass first decreases with increase in the transition density and then again increases as transition density increases. This results in 'U' like shapes with a minimum at some transition density as observed from the figure. This behavior is in  contrast with all the other  hadronic Eos where the maximum mass  mostly decreases with the transition density.  The maximum mass obtained from this EoS based on the NL3$\rho\omega$ parametrization also exceeds the range obtained from the pulsar  PSR J0740+6620 for the entire range of the CSS parameters used. For the EoS FSU2R, the maximum mass at lower energy gaps lie well below the constraints imposed by PSR 0952-607, specially for the lower values of the speed of sound. For the BSR8 EoS, the maximum mass configurations satisfy the observational constraint of PSR J0752–607 across both low and high values of the transition density $\rho_{\text{tr}}$. However, to satisfy the mass measurement of the  pulsar PSR J0740+6620, the EoS requires relatively higher values of $\rho_{\text{tr}}$ combined with lower values of  $C_s^2$ in the high-density phase.
  For DD2 EoS, the maximum mass calculated from the model spans the constraints of mass from both the pulsars at $C_s^2$ =0.6 whereas at higher speeds of sound the calculated maximum mass exceeds that of the pulsar PSR J0740+6620. 
  
To provide a more general perspective beyond individual models, we now synthesize the trends across the three physically viable EoSs: DD2, FSU2R, and BSR8 shown in the Fig.~\ref{fig:mr_psr_css_constraints} and in Table.\ref{tab:css_density_ranges}. This approach highlights the CSS parameter regions favored by observations without excessive dependence on specific hadronic inputs. Across all EoSs, increasing $C_s^2$ from 0.60 to 0.90 shifts the allowed $\Delta \varepsilon$ to higher values. A stiffer quark matter EoS allows for stronger phase transitions while maintaining a high maximum mass. The sensitivity of the constraints to $C_s^2$ is more pronounced for PSR~J0740+6620, which demands larger $\Delta \varepsilon$. At low $C_s^2$, increasing $\rho_{\text{tr}}$ does not necessarily lead to higher $\Delta \varepsilon$ values; instead, a non-monotonic pattern appears, particularly for  PSR J0740+6620.
When the phase transition happens at very high density (\(\rho_{\text{tr}}\)), the quark matter core occupies only a small volume. As a result, variations in the energy density jump \(\Delta \varepsilon\) have limited effect on the global properties (such as the maximum mass), especially when the quark matter is already stiff (\(C_s^2 \geq 0.75\)).
 PSR~J0740+6620 generally allows broader and higher values of the energy density gap $\Delta \varepsilon$, especially at moderate-to-high values of $C_s^2$, suggesting that a strong phase transition is needed to support its structure. In contrast, PSR~J0952--0607 imposes tighter constraints at high $\rho_{\text{tr}}$ and lower $C_s^2$ due to its higher mass, thereby excluding large $\Delta \varepsilon$ in that regime.

\begin{figure*}[htp] 
    \centering
    \includegraphics[width=0.30\textwidth]{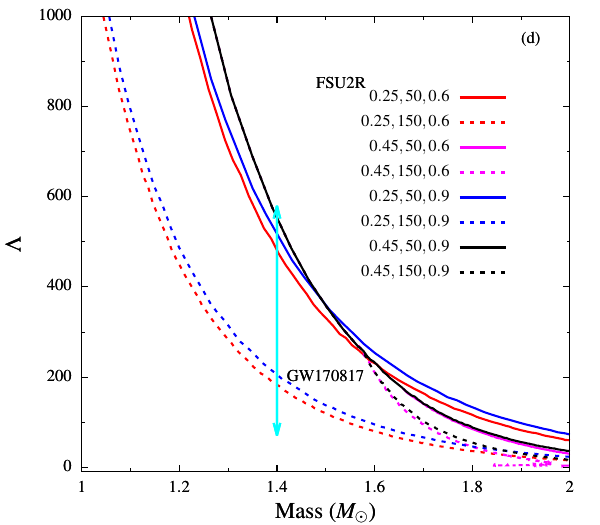}
    \includegraphics[width=0.30\textwidth]{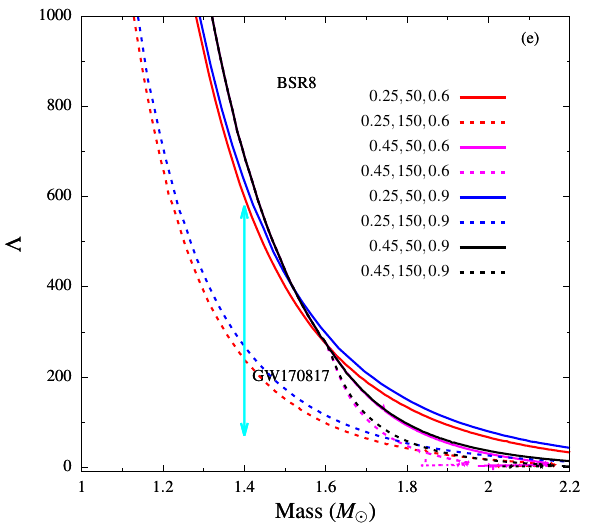}\includegraphics[width=0.30\textwidth]{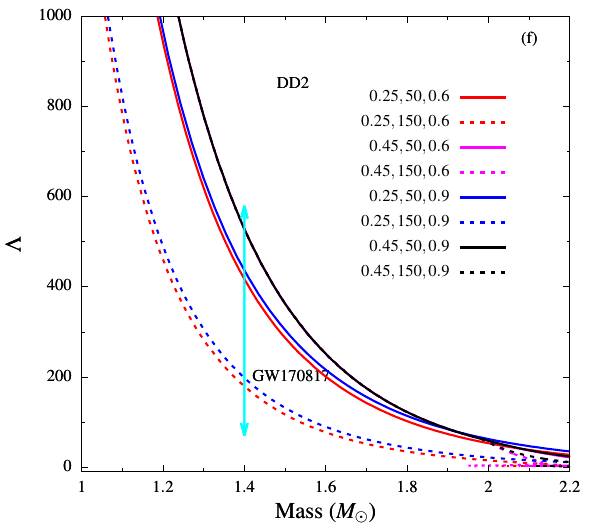}
    \caption{
Dimensionless tidal deformability $\Lambda$ as a function of mass for the hybrid star configurations corresponding to three physically viable models (FSU2R, BSR8 and DD2) . Each legend represents the CSS parameters: $\rho_{\text{tr}}$, $\Delta \varepsilon$, and $C_s^2$. The arrow indicates the $\Lambda_{1.4}$ constraint from the GW170817 event.}   
    \label{fig:tidal}
\end{figure*}


\begin{table}[!ht]
\centering
\begin{minipage}{0.30\textwidth}
\centering
\caption{DD2}
\setlength{\tabcolsep}{2pt}
\renewcommand{\arraystretch}{1.2}
\begin{tabular}{cccc}
\hline
\hline
$\rho_{\text{tr}}$ & $C_s^2$ & $\Delta \varepsilon$ (J0740) & $\Delta \varepsilon$ (J0952) \\
(fm$^{-3}$) & & (MeV fm$^{-3}$) & (MeV fm$^{-3}$) \\
\hline
0.25 & 0.60 & 160--210 & 50--140 \\
0.30 & 0.60 & 140--200 & 50--130 \\
0.35 & 0.60 & 130--190 & 50--110 \\
0.40 & 0.60 & 130--200 & 50--110 \\
0.45 & 0.60 & 140--230 & 50--120 \\
0.25 & 0.75 & 230--290 & 100--210 \\
0.30 & 0.75 & 200--280 & 70--190 \\
0.35 & 0.75 & 190--260 & 50--170 \\
0.40 & 0.75 & 190--260 & 50--160 \\
0.45 & 0.75 & 190--280 & 50--160 \\
0.25 & 0.90 & 280--300 & 140--260 \\
0.30 & 0.90 & 260--300 & 110--250 \\
0.35 & 0.90 & 250--300 & 80--220 \\
0.40 & 0.90 & 250--300 & 60--210 \\
0.45 & 0.90 & 250--300 & 50--200 \\
\hline
\hline
\end{tabular}
\end{minipage}
\hfill
\begin{minipage}{0.32\textwidth}
\centering
\caption{FSU2R}
\setlength{\tabcolsep}{2pt}
\renewcommand{\arraystretch}{1.2}
\begin{tabular}{cccc}
\hline
\hline
$\rho_{\text{tr}}$ & $C_s^2$ & $\Delta \varepsilon$ (J0740) & $\Delta \varepsilon$ (J0952) \\
(fm$^{-3}$) & & (MeV fm$^{-3}$) & (MeV fm$^{-3}$) \\
\hline
0.25 & 0.60 & 150--200 & 50--140 \\
0.30 & 0.60 & 120--175 & 50--120 \\
0.35 & 0.60 & 100--150 & 50--90 \\
0.40 & 0.60 & 75--140 & 50--70 \\
0.45 & 0.60 & 60--130 & -- \\
0.25 & 0.75 & 225--275 & 90--200 \\
0.30 & 0.75 & 200--250 & 60--175 \\
0.35 & 0.75 & 175--225 & 50--150 \\
0.40 & 0.75 & 140--200 & 50--130 \\
0.45 & 0.75 & 120--200 & 50--100 \\
0.25 & 0.90 & 275--325 & 130--250 \\
0.30 & 0.90 & 250--325 & 100--225 \\
0.35 & 0.90 & 225--300 & 60--200 \\
0.40 & 0.90 & 200--275 & 50--175 \\
0.45 & 0.90 & 175--250 & 50--150 \\
\hline
\hline

\end{tabular}
\end{minipage}
\begin{minipage}{0.32\textwidth}
\centering
\caption{BSR8}
\setlength{\tabcolsep}{2pt}
\renewcommand{\arraystretch}{1.2}
\begin{tabular}{cccc}
\hline
\hline
$\rho_{\text{tr}}$ & $C_s^2$ & $\Delta \varepsilon$ (J0740) & $\Delta \varepsilon$ (J0952) \\
(fm$^{-3}$) & & (MeV fm$^{-3}$) & (MeV fm$^{-3}$) \\
\hline
0.25 & 0.60 & 150--150 & 50--140 \\
0.30 & 0.60 & 120--150 & 50--125 \\
0.35 & 0.60 & 100--150 & 50--90  \\
0.40 & 0.60 &  70--140 & 50--60  \\
0.45 & 0.60 &  60--125 & --      \\
0.25 & 0.75 & --       & 90--150 \\
0.30 & 0.75 & --       & 50--150 \\
0.35 & 0.75 & --       & 50--150 \\
0.40 & 0.75 & 130--150 & 50--125 \\
0.45 & 0.75 & 120--150 & 50--100 \\
0.25 & 0.90 & --       & 130--150 \\
0.30 & 0.90 & --       & 90--150  \\
0.35 & 0.90 & --       & 60--150  \\
0.40 & 0.90 & --       & 50--150  \\
0.45 & 0.90 & --       & 50--150  \\

\hline
\hline

\end{tabular}
\end{minipage}
\label{tab:css_density_ranges}
\end{table}


\subsection*{Constraints on CSS parameters from the tidal deformability}
We investigate the impact of CSS model parameters on the tidal deformability 
\cite{Hinderer:2007mb,Flanagan:2007ix} for the selected hadronic models. In Fig.~\ref{fig:tidal}, we present the dimensionless tidal deformability ($\Lambda$) as a function of the neutron star mass for various parameter combinations. These results are particularly relevant in light of the gravitational wave constraints from the binary neutron star merger event GW170817 \cite{LIGOScientific:2018cki}. The tidal deformability of a 1.4 $M_{\odot}$ neutron star , based on the gravitational-wave event GW170817, is estimated to lie within the range 70–580.
In our analysis, we explore two representative values of the transition density, namely 0.25 and 0.45~$\text{fm}^{-3}$, and choose energy density gaps of 50 and 150~$\text{MeV~fm}^{-3}$. The squared speed of sound in the quark phase, $C_s^2$, is taken as 0.6 and 0.9. While $ \Lambda $ provides a valuable constraint on the hybrid star EoS, our analysis primarily focuses on as supporting check and the interplay between $\Lambda$ and the phase transition parameters within the context of physically viable models ( FSU2R, BSR8 and DD2 ). These models are selected based on their consistency with nuclear physics constraints and observed neutron star properties.

In Fig.~\ref{fig:tidal}, we analyze the tidal deformability predictions for the physically viable hadronic EOSs: FSU2R, DD2, and BSR8. For the FSU2R and DD2 models, we find that the entire range of phase transition parameters explored lies within the observationally allowed region for $\Lambda_{1.4}$, indicating robust compatibility with the GW170817 constraints. In the case of BSR8, although the agreement is somewhat more sensitive to the choice of parameters, we observe that lower transition densities and moderate energy density gaps tend to yield values of $\Lambda_{1.4}$ that better match the observational limits. 
Overall, our results show that the tidal deformability is sensitive to the details of the phase transition in the CSS model. The interplay between the transition density, the energy density discontinuity, and the speed of sound in the quark phase plays a crucial role in determining whether a given EOS is consistent with gravitational wave observations. These findings can help narrow down the viable parameter space for hybrid stars.
This behavior is consistent with well-established results in the literature, which have shown that $\Lambda$ is particularly sensitive to the nature of the hadron-quark phase transition within the CSS framework (\cite{Lenzi:2012xz,Pereira:2017rmp,Parisi:2020qfs,Lugones:2021bkm,Lenzi:2023nox,Mariani:2024gqi}). Our analysis reinforces this known dependence, further confirming that the transition density and energy density gap are key parameters in determining compatibility with tidal deformability constraints.

\section{Summary  and conclusion }\label{sec:conclusion}

\color{black} 
A reanalysis of PSR J0740+6620 and PSR J0030+0451 data tests the consistency of various nuclear EoS models.
Scenarios A, B, and C, based on different surface temperature models for PSR J0030+0451 along with PSRJ0740+6620, yield notably different mass-radius estimates. 
In light of the Bayesian evidence from \cite{Vinciguerra:2023qxq}, our analysis prioritizes Scenarios C and B, while Scenario A is included solely for completeness inspite of  its strong statistical disfavor.
In this work, we  have critically examined the hybrid star scenario in light of recent NICER measurements of PSR J0740+6620 and PSR J0030+0451. Motivated by the need to reconcile these observations with a theoretically consistent EoS,
our analysis demonstrate how the NICER mass-radius measurements impose constraints on the CSS model parameters. Prediction from the  CSS  model depends on the hadronic EoS with which it is combined, for that we have chosen six different hadronic EoS (using RMF models ) depending on the interaction (stiffness) term in the RMF Lagrangian.
Our results demonstrate that hybrid star configurations constructed using physically consistent hadronic EoSs like DD2, FSU2R, and BSR8 can successfully satisfy stringent observational constraints. This supports the viability of hybrid stars as realistic candidates for the internal composition of massive neutron stars.
To sharpen our analysis, we focused on two representative EoSs: FSU2R (density-independent) and DD2 (density-dependent), both consistent with empirical nuclear matter constraints. 
For the EoS based on the FSU2R parametrization, results across different $C_s^2$ values are similar at low energy density gaps ($\Delta \varepsilon$), but diverge significantly for $\Delta \varepsilon > 100\, \text{MeV~fm}^{-3}$, with the 2$M_{\odot}$ constraint disfavoring large gaps across all $\rho_{\text{tr}}$. Higher $C_s^2$ values (0.75, 0.9) show better agreement with observational constraints across categories A, B, and C.  For the next hadronic EOS, the density-dependent DD2, our model results meet the criteria of Category~B to a much lesser extent as 
 compared to FSU2R, up to $\Delta\varepsilon = 100\,\mathrm{MeV\,fm^{-3}}$
Each NICER scenario preferentially selects different regions of the hybrid EoS parameter space, reflecting their sensitivity to the microphysical properties of dense matter. While Scenario B favors lower transition densities and larger energy density gaps associated with more compact stars, Scenario C supports stiffer EoSs with delayed phase transitions, consistent with larger radii and extended hybrid star structures.
We have also constrained the CSS parameters using observational data from PSR J0740+6620 and PSR J0952-0607 by computing the maximum mass for these parameters. While PSRJ0740+6620 permits broader $\Delta\varepsilon$ ranges, especially at high $C_s^2$, PSRJ0952--0607 imposes tighter upper limits, particularly at high $\rho_{\text{tr}}$ and lower sound speeds.
Our analysis shows that the tidal deformability of neutron stars is highly sensitive to the CSS model parameters; with lower transition densities, higher energy gaps and lower quark phase sound speeds generally yielding better agreement with GW170817 constraints, with FSU2R and DD2 EOSs being more  compatible with these limits.

\textbf{Comparison with Previous Work:} Our study expands upon prior works such as those by \cite{Li:2024sft} by systematically incorporating all NICER scenarios and placing them within a broader CSS parameter study. While past analysis  have often considered a fixed hadron matter EoS, our study presents a more flexible and general hybrid star framework constrained by updated observational limits.

\textbf{Physical Implications and Outlook:} Our results reinforce the idea that hybrid stars---with realistic hadronic and quark phase properties---are viable candidates for describing compact stars. Notably, the possibility of hybridization in even canonical-mass stars at low $\rho_{\text{tr}}$ broadens the range of potential observational signatures. The stiffness of the hadronic EoS, the location of the phase transition, and the sound speed in quark matter all play decisive roles in determining the global structure and observable features of hybrid stars. Continued refinement of observational constraints on mass, radius, and tidal deformability will play a critical role in narrowing down the viable parameter space and improving our understanding of the QCD phase transition in astrophysical environments.

\color{black}

\bibliography{hs_css.bib}

\begin{thebibliography}{}
\expandafter\ifx\csname natexlab\endcsname\relax\def\natexlab#1{#1}\fi
\providecommand{\url}[1]{\href{#1}{#1}}
\providecommand{\dodoi}[1]{doi:~\href{http://doi.org/#1}{\nolinkurl{#1}}}
\providecommand{\doeprint}[1]{\href{http://ascl.net/#1}{\nolinkurl{http://ascl.net/#1}}}
\providecommand{\doarXiv}[1]{\href{https://arxiv.org/abs/#1}{\nolinkurl{https://arxiv.org/abs/#1}}}

\bibitem[{Abbott {et~al.}(2017)}]{LIGOScientific:2017vwq}
Abbott, B.~P., {et~al.} 2017, Phys. Rev. Lett., 119, 161101,
  \dodoi{10.1103/PhysRevLett.119.161101}

\bibitem[{Abbott {et~al.}(2018)}]{LIGOScientific:2018cki}
---. 2018, Phys. Rev. Lett., 121, 161101,
  \dodoi{10.1103/PhysRevLett.121.161101}

\bibitem[{Agrawal(2010)}]{Agrawal:2010wg}
Agrawal, B.~K. 2010, Phys. Rev. C, 81, 034323,
  \dodoi{10.1103/PhysRevC.81.034323}

\bibitem[{Alford {et~al.}(2013)Alford, Han, \& Prakash}]{PhysRevD.88.083013}
Alford, M.~G., Han, S., \& Prakash, M. 2013, Phys. Rev. D, 88, 083013,
  \dodoi{10.1103/PhysRevD.88.083013}

\bibitem[{Bhattacharyya {et~al.}(2010)Bhattacharyya, Mishustin, \&
  Greiner}]{Bhattacharyya:2009fg}
Bhattacharyya, A., Mishustin, I.~N., \& Greiner, W. 2010, J. Phys. G, 37,
  025201, \dodoi{10.1088/0954-3899/37/2/025201}

\bibitem[{Blaschke \& Cierniak(2021)}]{Blaschke:2020vuy}
Blaschke, D., \& Cierniak, M. 2021, Astron. Nachr., 342, 227,
  \dodoi{10.1002/asna.202113909}

\bibitem[{Bozzola {et~al.}(2019)Bozzola, Espino, Lewin, \&
  Paschalidis}]{Bozzola:2019tit}
Bozzola, G., Espino, P.~L., Lewin, C.~D., \& Paschalidis, V. 2019, Eur. Phys.
  J. A, 55, 149, \dodoi{10.1140/epja/i2019-12831-2}

\bibitem[{Brandes \& Weise(2025)}]{Brandes:2024wpq}
Brandes, L., \& Weise, W. 2025, Phys. Rev. D, 111, 034005,
  \dodoi{10.1103/PhysRevD.111.034005}

\bibitem[{Christian {et~al.}(2019)Christian, Zacchi, \&
  Schaffner-Bielich}]{Christian:2018jyd}
Christian, J.-E., Zacchi, A., \& Schaffner-Bielich, J. 2019, Phys. Rev. D, 99,
  023009, \dodoi{10.1103/PhysRevD.99.023009}

\bibitem[{Dhiman {et~al.}(2007)Dhiman, Kumar, \& Agrawal}]{Dhiman:2007ck}
Dhiman, S.~K., Kumar, R., \& Agrawal, B.~K. 2007, Phys. Rev. C, 76, 045801,
  \dodoi{10.1103/PhysRevC.76.045801}

\bibitem[{Drischler {et~al.}(2021)Drischler, Han, Lattimer, Prakash, Reddy, \&
  Zhao}]{Drischler:2020fvz}
Drischler, C., Han, S., Lattimer, J.~M., {et~al.} 2021, Phys. Rev. C, 103,
  045808, \dodoi{10.1103/PhysRevC.103.045808}

\bibitem[{Dutra {et~al.}(2014)Dutra, Louren\c{c}o, Avancini, Carlson, Delfino,
  Menezes, Provid\^encia, Typel, \& Stone}]{Dutra:2014qga}
Dutra, M., Louren\c{c}o, O., Avancini, S.~S., {et~al.} 2014, Phys. Rev. C, 90,
  055203, \dodoi{10.1103/PhysRevC.90.055203}

\bibitem[{Ferreira {et~al.}(2020)Ferreira, C\^amara~Pereira, \&
  Provid\^encia}]{Ferreira:2020evu}
Ferreira, M., C\^amara~Pereira, R., \& Provid\^encia, C. 2020, Phys. Rev. D,
  101, 123030, \dodoi{10.1103/PhysRevD.101.123030}

\bibitem[{Flanagan \& Hinderer(2008)}]{Flanagan:2007ix}
Flanagan, E.~E., \& Hinderer, T. 2008, Phys. Rev. D, 77, 021502,
  \dodoi{10.1103/PhysRevD.77.021502}

\bibitem[{Geng {et~al.}(2003)Geng, Toki, Sugimoto, \& Meng}]{Geng:2003pk}
Geng, L.-s., Toki, H., Sugimoto, S., \& Meng, J. 2003, Prog. Theor. Phys., 110,
  921, \dodoi{10.1143/PTP.110.921}

\bibitem[{Glendenning(2000)}]{Glendenning:1997wn}
Glendenning, N.~K. 2000, {Compact stars: Nuclear physics, particle physics, and
  general relativity} (Springer-Verlag, New York)

\bibitem[{Glendenning \& Moszkowski(1991)}]{Glendenning:1991es}
Glendenning, N.~K., \& Moszkowski, S.~A. 1991, Phys. Rev. Lett., 67, 2414,
  \dodoi{10.1103/PhysRevLett.67.2414}

\bibitem[{Gomes {et~al.}(2019)Gomes, Char, \& Schramm}]{Gomes:2018eiv}
Gomes, R.~O., Char, P., \& Schramm, S. 2019, Astrophys. J., 877, 139,
  \dodoi{10.3847/1538-4357/ab1751}

\bibitem[{Han {et~al.}(2019)Han, Mamun, Lalit, Constantinou, \&
  Prakash}]{Han:2019bub}
Han, S., Mamun, M. A.~A., Lalit, S., Constantinou, C., \& Prakash, M. 2019,
  Phys. Rev. D, 100, 103022, \dodoi{10.1103/PhysRevD.100.103022}

\bibitem[{Hinderer(2008)}]{Hinderer:2007mb}
Hinderer, T. 2008, Astrophys. J., 677, 1216, \dodoi{10.1086/533487}

\bibitem[{Horowitz \& Piekarewicz(2001)}]{Horowitz:2000xj}
Horowitz, C.~J., \& Piekarewicz, J. 2001, Phys. Rev. Lett., 86, 5647,
  \dodoi{10.1103/PhysRevLett.86.5647}

\bibitem[{Laskos-Patkos \& Moustakidis(2023)}]{Laskos-Patkos:2023cts}
Laskos-Patkos, P., \& Moustakidis, C.~C. 2023, Phys. Rev. D, 107, 123023,
  \dodoi{10.1103/PhysRevD.107.123023}

\bibitem[{Lenzi \& Lugones(2012)}]{Lenzi:2012xz}
Lenzi, C.~H., \& Lugones, G. 2012, Astrophys. J., 759, 57,
  \dodoi{10.1088/0004-637X/759/1/57}

\bibitem[{Lenzi {et~al.}(2023)Lenzi, Lugones, \& Vasquez}]{Lenzi:2023nox}
Lenzi, C.~H., Lugones, G., \& Vasquez, C. 2023, Phys. Rev. D, 107, 083025,
  \dodoi{10.1103/PhysRevD.107.083025}

\bibitem[{Li {et~al.}(2025)Li, Sedrakian, \& Alford}]{Li:2024sft}
Li, J.~J., Sedrakian, A., \& Alford, M. 2025, JCAP, 02, 002,
  \dodoi{10.1088/1475-7516/2025/02/002}

\bibitem[{Liu {et~al.}(2022)Liu, Xu, \& Chu}]{Liu:2022mje}
Liu, H., Xu, J., \& Chu, P.-C. 2022, Phys. Rev. D, 105, 043015,
  \dodoi{10.1103/PhysRevD.105.043015}

\bibitem[{Lugones \& Grunfeld(2021)}]{Lugones:2021tee}
Lugones, G., \& Grunfeld, A.~G. 2021, Phys. Rev. D, 104, L101301,
  \dodoi{10.1103/PhysRevD.104.L101301}

\bibitem[{Lugones {et~al.}(2023)Lugones, Mariani, \&
  Ranea-Sandoval}]{Lugones:2021bkm}
Lugones, G., Mariani, M., \& Ranea-Sandoval, I.~F. 2023, JCAP, 03, 028,
  \dodoi{10.1088/1475-7516/2023/03/028}

\bibitem[{Mariani {et~al.}(2024)Mariani, Ranea-Sandoval, Lugones, \&
  Orsaria}]{Mariani:2024gqi}
Mariani, M., Ranea-Sandoval, I.~F., Lugones, G., \& Orsaria, M.~G. 2024, Phys.
  Rev. D, 110, 043026, \dodoi{10.1103/PhysRevD.110.043026}

\bibitem[{Maslov {et~al.}(2019)Maslov, Yasutake, Ayriyan, Blaschke, Grigorian,
  Maruyama, Tatsumi, \& Voskresensky}]{Maslov:2018ghi}
Maslov, K., Yasutake, N., Ayriyan, A., {et~al.} 2019, Phys. Rev. C, 100,
  025802, \dodoi{10.1103/PhysRevC.100.025802}

\bibitem[{Miller {et~al.}(2019)}]{Miller:2019cac}
Miller, M.~C., {et~al.} 2019, Astrophys. J. Lett., 887, L24,
  \dodoi{10.3847/2041-8213/ab50c5}

\bibitem[{Miller {et~al.}(2021)}]{Miller:2021qha}
---. 2021, Astrophys. J. Lett., 918, L28, \dodoi{10.3847/2041-8213/ac089b}

\bibitem[{Montana {et~al.}(2019)Montana, Tolos, Hanauske, \&
  Rezzolla}]{Montana:2018bkb}
Montana, G., Tolos, L., Hanauske, M., \& Rezzolla, L. 2019, Phys. Rev. D, 99,
  103009, \dodoi{10.1103/PhysRevD.99.103009}

\bibitem[{Most {et~al.}(2019)Most, Papenfort, Dexheimer, Hanauske, Schramm,
  St\"ocker, \& Rezzolla}]{Most:2018eaw}
Most, E.~R., Papenfort, L.~J., Dexheimer, V., {et~al.} 2019, Phys. Rev. Lett.,
  122, 061101, \dodoi{10.1103/PhysRevLett.122.061101}

\bibitem[{Pal {et~al.}(2025)Pal, Podder, \& Chaudhuri}]{Pal:2025skz}
Pal, S., Podder, S., \& Chaudhuri, G. 2025, Astrophys. J., 983, 24,
  \dodoi{10.3847/1538-4357/adbc6b}

\bibitem[{Parisi {et~al.}(2020)Parisi, V\'asquez~Flores, Lenzi, Chen, \&
  Lugones}]{Parisi:2020qfs}
Parisi, A., V\'asquez~Flores, C., Lenzi, C.~H., Chen, C.-S., \& Lugones, G.
  2020, \dodoi{10.1088/1475-7516/2021/06/042}

\bibitem[{Pereira {et~al.}(2018)Pereira, Flores, \& Lugones}]{Pereira:2017rmp}
Pereira, J.~P., Flores, C.~V., \& Lugones, G. 2018, Astrophys. J., 860, 12,
  \dodoi{10.3847/1538-4357/aabfbf}

\bibitem[{Riley {et~al.}(2019)}]{Riley:2019yda}
Riley, T.~E., {et~al.} 2019, Astrophys. J. Lett., 887, L21,
  \dodoi{10.3847/2041-8213/ab481c}

\bibitem[{Riley {et~al.}(2021)}]{Riley:2021pdl}
---. 2021, Astrophys. J. Lett., 918, L27, \dodoi{10.3847/2041-8213/ac0a81}

\bibitem[{Salmi {et~al.}(2022)}]{Salmi:2022cgy}
Salmi, T., {et~al.} 2022, Astrophys. J., 941, 150,
  \dodoi{10.3847/1538-4357/ac983d}

\bibitem[{Sun {et~al.}(2024)Sun, Bhattiprolu, \& Lattimer}]{Sun:2023xkg}
Sun, B., Bhattiprolu, S., \& Lattimer, J.~M. 2024, Phys. Rev. C, 109, 055801,
  \dodoi{10.1103/PhysRevC.109.055801}

\bibitem[{Tolos {et~al.}(2017)Tolos, Centelles, \& Ramos}]{Tolos:2016hhl}
Tolos, L., Centelles, M., \& Ramos, A. 2017, Astrophys. J., 834, 3,
  \dodoi{10.3847/1538-4357/834/1/3}

\bibitem[{Tsaloukidis {et~al.}(2023)Tsaloukidis, Koliogiannis, Kanakis-Pegios,
  \& Moustakidis}]{Tsaloukidis:2022rus}
Tsaloukidis, L., Koliogiannis, P.~S., Kanakis-Pegios, A., \& Moustakidis, C.~C.
  2023, Phys. Rev. D, 107, 023012, \dodoi{10.1103/PhysRevD.107.023012}

\bibitem[{Typel {et~al.}(2010)Typel, Ropke, Klahn, Blaschke, \&
  Wolter}]{Typel:2009sy}
Typel, S., Ropke, G., Klahn, T., Blaschke, D., \& Wolter, H.~H. 2010, Phys.
  Rev. C, 81, 015803, \dodoi{10.1103/PhysRevC.81.015803}

\bibitem[{Vinciguerra {et~al.}(2024)}]{Vinciguerra:2023qxq}
Vinciguerra, S., {et~al.} 2024, Astrophys. J., 961, 62,
  \dodoi{10.3847/1538-4357/acfb83}

\bibitem[{Weissenborn {et~al.}(2011)Weissenborn, Sagert, Pagliara, Hempel, \&
  Schaffner-Bielich}]{Weissenborn:2011qu}
Weissenborn, S., Sagert, I., Pagliara, G., Hempel, M., \& Schaffner-Bielich, J.
  2011, Astrophys. J. Lett., 740, L14, \dodoi{10.1088/2041-8205/740/1/L14}

\bibitem[{Xia {et~al.}(2019)Xia, Maruyama, Yasutake, \& Tatsumi}]{Xia:2019pnq}
Xia, C.-J., Maruyama, T., Yasutake, N., \& Tatsumi, T. 2019, Phys. Rev. D, 99,
  103017, \dodoi{10.1103/PhysRevD.99.103017}

\bibitem[{Zha {et~al.}(2020)Zha, O'Connor, Chu, Lin, \& Couch}]{Zha:2020gjw}
Zha, S., O'Connor, E.~P., Chu, M.-c., Lin, L.-M., \& Couch, S.~M. 2020, Phys.
  Rev. Lett., 125, 051102, \dodoi{10.1103/PhysRevLett.127.219901}

\end{thebibliography}

\end{document}